\documentclass[a4paper,10pt]{article}

\usepackage[round,authoryear,sort]{natbib}
\bibpunct{(}{)}{,}{a}{,}{,~}
\newcommand{\forget}[1]{}

\title{The theory of the rise of sap in trees: some historical and conceptual remarks}

\author{ Harvey R. Brown\\
Faculty of Philosophy, University of Oxford\\ Radcliffe Humanities, Radcliffe Observatory Quarter\\Woodstock Road, Oxford OX2
6GG, U.K.\\{\em harvey.brown@philosophy.ox.ac.uk}\\}

\date{}

\begin{document}

\maketitle

\begin{abstract}
The ability of trees to suck water from roots to leaves, sometimes to heights of over a hundred meters, is remarkable given the absence of any mechanical pump. This study deals with a number of issues, of both an historical and conceptual nature, in the orthodox ``Cohesion-Tension''  theory of the ascent of sap in trees. The theory relies chiefly on the exceptional cohesive and adhesive properties of water, the structural properties of trees, and the role of evaporation (``transpiration'') from leaves. But it is not the whole story. Plant scientists have been aware since the inception of the theory in the late 19th century that further processes are at work in order to ÒprimeÓ the trees, the main such process --  growth itself -- being so obvious to them that it is often omitted from the story. Other factors depend largely on the type of tree, and are not always fully understood. For physicists, in particular, it may be helpful to see the fuller picture, which is what this study attempts to provide in non-technical terms.\footnote{This paper is an updated and reformatted version of Brown (2013), with a number of corrections and additions (especially in section 6.3).}
\end{abstract}

``\textit{There are therefore agents in Nature able to make the particles of bodies stick together by very strong attractions. And it is the business of experimental philosophy to find them out.}" Isaac Newton\footnote{Query 31 in the \textit{Opticks}; Newton (1979), p. 394.}
\bigskip

``\textit{To believe that columns of water should hang in the tracheals like solid bodies, and should, like them, transmit downwards the pull exerted on them at their upper ends by the transpiring leaves, is to some of us equivalent to believing in ropes of sand.}'' Francis Darwin\footnote{Darwin \textit{et al.} (1896), p. 635.}
\bigskip

``\textit{Water is unique in its importance and its properties. No other substance has been the subject of so much study and speculation, nor has any been harder to understand at a molecular level.}'' J.S. Rowlinson\footnote{Rowlinson (2005), p. 262.}

\section{Spat over sap}

In 2004, a brief, vehement letter appeared in the botany journal \textit{New Phytologist}, signed by 45 scientists from seven countries.\footnote{Angeles \textit{et al.} (2004).} It repudiated a recent review paper in the same journal\footnote{Zimmermann \textit{et al.} (2004).} which had attempted to show that standard arguments in favour of the so-called Cohesion-Tension theory, the conventional wisdom concerning the remarkable mechanism behind the ascent of sap in trees, are ``completely misleading". The review paper's leading author was Ulrich Zimmermann. This is not without irony; one of the leading figures in the twentieth century development of the lambasted Cohesion-Tension Theory 
was Martin H. Zimmermann.\footnote{Martin Zimmermann, who died in 1984, was Charles Bullard Professor at Harvard University and Director of the Harvard Forest from 1970 until his death, and the author of an authoritative 1983 textbook on the ascent of sap. Ulrich Zimmermann is currently Senior Professor at the Biocenter of the University of W\"{u}rzburg, Germany.}
The 45-author rebuttal finished with the sentence: ``We wish the readers of \textit{New Phytologist} to know that the Cohesion-Tension theory is widely supported as the only theory consistent with the preponderance of data on water transport in plants." The letter was followed by an editorial in which the the Editor-in-Chief of  \textit{New Phytologist} felt the need to state the obvious  --- that views expressed in any review or paper appearing in the journal belong to the authors alone.

The dispute should probably not be seen as deeply divisive within the plant science community; the sheer number of authors in the letter attests to this. Not that all aspects of the ascent of sap are fully understood within the Cohesion Tension (CT) theory, as we shall see. But what makes this scientific spat stand out from so many others are not so much the details under dispute nor the strident, sometimes scathing tone of the debate. There is the sheer wonder of the subject matter itself. For hundreds of millions of years, plants have existed which propel water vertically from their roots through stems into their leaves. Wood itself evolved around 400 million years ago in all likelihood primarily to improve water conductance in plants, rather than to provide mechanical support.\footnote{See Gerrienne \textit{et al.} (2011).}   The magnitude of the resulting hydraulic marvel in trees is brought into focus when one considers the height of the tallest trees, their age, and the fact that the main suction mechanism is unlike anything we experience in daily life. All around us are found living structures that perform the feat of internally transporting water, or rather water-based solutions (sap) -- in some cases hundreds of litres in a day -- against the force of gravity, using no metabolic pumps and no moving parts apart from stomata (pores) in the leaves. For some trees, the process can go on for hundreds, even thousands of years, despite all the abuse Nature hurls at them in the form of storms, floods, frost, fire, and drought. Not to mention attacks from bacteria and viruses, infestations of fungi and insects, and larger predators grazing on seeds and shoots.\footnote{For a variety of reasons, large old trees are undergoing a global decline and may be imperiled. See  Lindenmayer \textit{et al.} (2012).}

The tallest trees on Earth today are coastal redwoods, \textit{Sequoia sempervirens}, of northern California, the three supremos of which were only discovered in 2006. The tallest is over 115 meters. Given that only 5\% of the original redwood forests have survived logging, it is hard to believe that taller specimens have not existed.\footnote{See Koch \textit{et al}. (2006), who note that for California redwoods, summer fog is an important supply of water, and that direct absorption by the leaves may supplement the uptake from root capture of fog drip.}  Redwoods are not the only skyscrapers of the plant world, however. Douglas-fir, \textit{Pseudotsuga menziesii}, from Washington and British Columbia, and mountain ash, \textit{Eucalyptus regnans}, from Tasmania and Victoria, Australia, can reach similar heights.\footnote{Richard McArdle, a former chief of the U.S. Forest Service, estimated a Douglas-fir in Washington to be 120m in 1924; see Carder (1995). The tallest Australian eucalypt today is 100m, and there is strong anecdotal evidence of taller ones in the past. The tallest known Sitka spruce \textit{Picea sitchensis} in California has a height of nearly 97 m.} As for longevity, redwoods can live for longer than a millennium\footnote{A redwood tree's annual rate of wood production increases with age for at least 1,500 years; see J. Bourne (2009).}, but they are far from establishing the record. A living bristlecone pine, \textit{Pinus longaeva}, in the inhospitable White-Inyo Mountains in California, is known to have germinated well over four thousand years ago.\footnote{It is noteworthy that the oldest trees can live on dry sites; as befits such conditions bristlecone pines are small, thick trees with highly reduced growth rates (see Kozlowski and Pallardy (1996) p. 299). Trees allegedly even older than the bristlecone pine are referred to in Suzuki and Grady (2004), Ch. 5, but such claims seem to be rare in the literature. Note that the bristlecone pine is a non-clonal tree; the oldest individual vegetatively cloned tree, which repeatedly sprouted from the same or newly cloned roots, was discovered in Sweden in 2008, a spruce aged over 9000 years.  Clonal colonies (multiple trees connected by a common root system) can be much older.} 

\section{The problem}

Any plant with leaves over a few meters above its roots would, it seems, have to be using a mechanism or mechanisms other than barometric pressure or capillarity in the stem to get water from the roots to the leaves. The barometric effect could in principle only account for a 10.4 meter rise. Capillary rise varies with the inverse of the vessel diameter; diameters of xylem conduits in tree trunks
vary from 10 to 20$\mu$m in conifers to approximately 100$\mu$m in broad-leaved woody species.\footnote{Xylem, a portion of the sapwood, consists of repeating hollow elements that come in two kinds: \textit{vessels} (as in broad-leaved trees) and the evolutionarily more primitive \textit{tracheids} (as in conifers). We shall not be concerned in the main with the differences between these kinds, and will generally refer to ``conduits" or ``elements" without discrimination. The hollow interior of the conduit is referred to as the \textit{lumen} (plural \textit{lumina}). A spectacular collection of electron microscope photographs of woody structure is found in Meylan and Butterfield (1972).} Even with the optimal 10$\mu$m, a sap rise of only 3m is achieved.\footnote{See, for example, Steudle (1995); we shall return to this issue in  section 6.} It is also widely held that positive pressure, known to be provided by the osmotic flow mechanism in the roots, is likewise too meager in the case of most tall trees. 

  The ascent of sap in trees has long intrigued natural philosophers, but it was not until the end of the 19th century that the basic elements of the mechanism were discovered (or so it is commonly said). Why did it take so long? The theoretical possibility that water has the ability to form hanging threads tens of meters long, which remain unbroken even when the threads are pulled -- not pushed -- upwards, took two hundred years to mature. And that it appeared in the botany literature so late in the day has surely to do with the fact that this tensile phenomenon is hidden inside trees, and not obvious anywhere else in Nature.\footnote{John Joly, one of the fathers of the CT theory, remarked (Darwin \textit{et al}. (1896) p. 658) that ``we do not know if hydrostatic tension has been detected anywhere else in nature''. It was realised early in the 20th century that the remarkable mechanism of spore ejection from fern sporangia also testifies to the high tensile strength of water; here the tensions may be several times greater than in typical trees. (See Tyree and Zimmerman (2002), p. 63.) In 1991 it was realised that octopus suckers can also generate tension in sea water; see Smith (1991). But the effects of cavitation -- a consequence of negative pressures in liquids, as we shall see -- had been studied by Lord Raleigh in the late 19th century, and ship builders were aware of effects of cavitation on propeller efficiency and integrity. The evidence for liquids in a tensile state in artificial conditions in fact goes back to 1661; see section 5.1 below.}  Even today, it seems, there are those in the plant science community who question whether the degree of ``negative pressure''\footnote{Negative pressure is defined as pressure less than atmospheric pressure, or the vapour pressure of water.} in sap postulated by the CT theory really exists in xylem.  There is no doubt, however, that the theory is the result of a great deal of hard scientific labour, and for the large majority of plant scientists it has  stood the test of time. It is a quiet triumph of modern science. 
  
But a non-plant scientist is likely, on first exposure, to find the theory incomplete. As we shall see, the cohesion mechanism tied to evaporation clearly requires `priming'; what is the mechanism for that? The \textit{short} answer -- growth itself -- is so obvious to modern plant scientists that it is usually omitted from the story. The \textit{long} answer is very intricate, depends on the type of tree, and still contains a number of real mysteries. This essay is written for non-specialists, so emphasis on the priming issue is greater than in standard accounts in the plant science literature. 
 
\section{Transpiration}

There is widespread appreciation of the unparalleled importance wood has had in the history of humanity as fuel and material for shelter, tools, ship-building, ornament, and so on,\footnote{The weight of wood used today as a raw material in the United States is still greater than the weight of all metals and plastics combined; see Shmulsky and Jones (2011), p. 443.} and it is reflected in the popular literature.\footnote{See, for example, Tudge (2006).} The importance of trees as anti-gravity devices may be less familiar, but it is more relevant than ever.

On a dry day, the water that evaporates, or ``transpires", from a tree's leaves is largely drawn from its roots.\footnote{The possibility of some evaporated water originating in storage compartments inside trees will be discussed at the end of this essay.} The release of water vapour into the atmosphere from tree foliage -- both through transpiration and evaporation of captured rainfall -- is one of the key factors in determining local climate. This is particularly so in the case of the rain forest, which ``behaves like a green ocean, transpiring water that rains upward, as though gravity were reversed.''\footnote{Suzuki and Grady (2004), Ch. 2. It is noteworthy that in this elegant and highly readable book on trees, built around the life cycle of a Douglas-fir, the authors present a confused account of the mechanism of sap rise,  ultimately regarding it as a ``mystery'' (Ch. 2).}  Globally, the amount of water transpired by plants in a year is roughly double that which flows in all the planet's rivers, adding 32 $\times 10^{3}$ billion tonnes of water vapour to the atmosphere\footnote{See Beerling and Franks (2010) and Steudle (1995). 
It has been widely held that the principal reason for this rate of transpiration is the fact that xylem sap in plants is very dilute -- large amounts have to be transpired so that sufficient quantities of soil nutriments can be accumulated in leaf cells (a notion that goes back at least to Stephen Hales in 1727; see Hales (1727), p. 6). However, since the 1990s, the idea that plant growth relies on transpiration has increasingly been called into question. It seems that the evolution of a leaf structure favouring high rates of photosynthesis -- and hence intake of CO$_2$ into the leaves through the stomata, the same apertures allowing for escape of water vapour --  has in most habitats had greater survival value than one conserving water. See Tanner and Beevers (1990), Kramer and Boyer (1995), p. 203, and Beerling and Franks (2010).} --- though evaporation from the oceans accounts for much more.
In recent years, the phenomenon of transpiration from forests has caught the attention of climate scientists concerned with the enhanced greenhouse effect associated with rising CO$_2$ emissions.
Transpiration cools, in precisely the way evaporation of sweat cools the body.\footnote{It is sometimes said that under a hot sun leaves would often perish in the absence of evaporative cooling, but not all plant scientists agree; see Kramer and Boyer (1995), p. 203.} But water vapour is a greenhouse gas, more potent than CO$_2$, so transpiration warms as well. The tricky business of evaluating whether cooling or warming wins out also depends on latitude and local background climate.  

Forests have a low albedo factor. Being dark, they reflect incoming solar radiation less than the ground underneath them would even if covered in vegetation, and so have a warming effect. This is particularly so for high latitude boreal forests (which constitute the largest biome on Earth) because for at least part of the year they may shade ground covered with snow, a strong reflector. In recent years, the startling possibility has been raised that boreal reforestation and afforestation -- the establishment of a forest on land not previously forested -- particularly in the coldest regions may ultimately exacerbate global warming.\footnote{During the cooling periods in glacial cycles over the last two million years, loss of boreal forest is believed to have provided a positive feedback in relation to glaciation. Conversely, 6000 years ago the expansion of boreal forests following the shrinkage of ice-sheets from the last ice age would have given rise to a positive feedback on warming. See Bonan (2008), p. 1445. In the case of lower latitude boreal forests, with less snow cover, the warming effect of low forest albedo may however be offset by cooling due to the release of organic vapours, which result in increasing cloud nuclei condensations and hence greater cloud albedo; see Spracklen \textit{et al.} (2008). A more recent study indeed provides evidence of a negative feedback at work in continental mid- and high-latitude environments: warming temperatures increase biogenic aerosol emissions, enhancing cloud albedo; see Paasonen \textit{et al.} (2013).} 
 In relation to the tropics, however, there is more consensus that reforestation and afforestation mitigate global warming. This is partly because the albedo effect is less of an issue, and again partly because higher temperatures and more plentiful soil moisture yield more water vapour, which in turn leads to the formation of more low-altitude clouds and precipitation. These processes have an overall cooling effect.\footnote{See Bonan (2008) and Betts (2011).}

The hydrological implications of changing large-scale forest-cover constitute an important and increasingly recognised ingredient in the understanding of global climate change.\footnote{In assessing the long-term role of forests in climate change, the back reaction of global warming on the forests must also be taken into consideration. For example, in a warmer world with less snow, the decrease of albedo linked to the masking of snow becomes less significant, and carbon storage in trees increases as a result of future CO$_2$ fertilization. On the other hand, if droughts become more frequent, the evaporative cooling of forests diminishes.(See Bala et al (2007) and especially the review by Bonan (2008). More recent studies are found in Arora \textit{et al.} (2011) and Pongratz \textit{et al.} (2011).) A drying trend in the Southern Hemisphere has led in the decade 2000-2009 to a global decrease in the amount of atmospheric carbon fixed by plants and accumulated as biomass. (See Zhao and Running (2010).) Very recently it has also been discovered that the advance of forests into Arctic tundra caused by rising temperatures releases carbon into the atmosphere from the soil through a process called positive priming. Such release of carbon may outweigh the carbon captured in growth of the forests, another way boreal afforestation may accelerate climate change. (See Hartley \textit{et al.} (2012).) Finally, higher temperatures and changing rainfall patterns also encourage damaging insect and fungal infestations; for example, high elevation pine -- including bristlecones -- in the West of the United States and Canada are increasingly vulnerable to mountain pine beetles and whitepine blister rust. (See Eilperin (2012) and Quinton (2012).)} It is all the more striking then that not only is the mechanism of sap rise in trees still subject to a degree of debate, but there is, to the author's knowledge, no extensive, up-to-date history of the relevant theory.\footnote{Historical sketches of varying lengths can be found in Darwin \textit{et al.} (1896), Copeland (1902), Dixon (1914),  Miller (1938), pp. 855-872, Kozlowski and Pallardy (1996) p. 259. Greenridge (1957), Pickard (1981), and especially the online essay by Richter and Cruiziat (2002).  It is noteworthy that in J. S. Rowlinson's admirable semi-historical 2002 book on the physics of cohesion, which contains fascinating detail of the long struggle to understand the cohesive properties of water, there is but a single brief mention of the rise of sap in trees; see Rowlinson (2002), p. 19.} One of the chief aims of the remarks below is to 
 encourage historians of science to undertake such a study.

\section{The orthodox theory and its origins}

\subsection{The CT theory}

The Cohesion-Tension theory, as it is standardly formulated, rests on a number of basic claims.\footnote{Succinct outlines of the theory are found in Holbrook \textit{et al.} (2002) and Cruiziat and Richter (2006). More detailed review papers are Greenidge (1957), Canny (1977), Steudle (2001); lengthier treatments are found in in the textbooks of Kramer and Boyer (1995), Kozlowski and Pallardy (1996), Tyree and Zimmerman (2002) (this being the second edition of the authoritative Zimmermann (1983)),  Ehlers and Goss (2003) and Nobel (2005). For a physicist, the technical review paper by Pickard (1981) may be the most informative and satisfying, though it is somewhat out of date, particularly in relation to mechanisms for refilling embolized xylem conduits. An excellent, up-to-date, somewhat less technical review is found in Sperry (2011). The literature is huge; these publications represent a very small selection; a useful list of key papers is found in Angeles \textit{et al.} (2004).} 

\textit{First}, that inside trees, sap forms a myriad of broken and more importantly \textit{unbroken} threads, stretching from the absorbing surfaces of the roots to the evaporation surfaces of the leaves, through the vascular structure of the xylem  in the stem. \textit{Second}, that transpiration principally from the leaves is the trigger of the driving force behind the ascent of sap.\footnote{Transpiration takes place through the bark, branches and twigs of trees, but by far the greatest amount is through leaves when there are any. Of that amount the greatest part is through the stomata, which open and close depending on a number of factors both within the leaf and without. The CO$_2$ absorbed from the atmosphere passes in the opposite direction through the open stomata; but for every CO$_2$ molecule used for the production of sugars, several hundred water molecules are released to the atmosphere. Trees are thirsty; see Holbrook \textit{et al.} (2002), p. 493. As Sperry (2011) puts it: ``The transpiration stream represents a river of water flowing in exchange for a relative trickle of carbon.''} Transpiration, or the removal of water molecules from the water-air interfaces (menisci) within the pores of certain mesophyll cell walls in the leaves\footnote{The exact site of the evaporation surfaces inside the leaves is still the subject of debate; see e.g. Kramer and Boyer (1995), p. 204.}, causes the interfaces to recede into the pores and change their shape (i.e. become more concave). The combination of forces of adhesion to the cell walls and surface tension on the interfaces --- capillarity, in short\footnote{A useful brief account of the physics of capillarity is found in Pickard (1981), section III; for more background, see Rowlinson (2002).} --- acts to restore the equilibrium shape of the menisci and thereby creates tension (negative pressure) in the sap in their vicinity. This tension in turn is transmitted to adjacent regions, including adjoining cell walls and cell protoplasts, then to the sap in the xylem, and eventually in the case of unbroken threads to the roots themselves.\footnote{We shall return to this process in more detail in section 4.3 below.} \textit{Third}, such increase in tension in the roots leads to greater passive absorption of water from the soil, so that water lost in transpiration in the foliage is replaced.\footnote{There is however often a lag between absorption of water in the roots and the onset of transpiration in the morning; the reasons for this will be touched on later. The speeds of the transpiration flow in coniferous tracheids are in the range 20-40 cm hr$^{-1}$; in the vessels of broad-leaved trees, a common speed is 5 m hr$^{-1}$, though 44 m hr$^{-1}$ has been recorded in an oak. See Canny (1977), p. 285, and Meinzer \textit{et al.} (2006).} \textit{Fourth}, the energy for the whole process ultimately comes from the sun, in overcoming the latent heat of evaporation of the water molecule -- the energy needed to break hydrogen bonds between water molecules at the menisci.\footnote{Direct solar radiation is only one factor; there is also radiation from the soil and surrounding objects, as well as heat being provided by the surrounding air. see Kramer and Boyer (1995) p. 207, and Ehlers and Goss (2003), p. 89.}  \textit{Fifth}, because of the remarkable cohesive nature of water, the rising columns of sap can mostly remain intact under such tension in the hours of transpiration. They are superheated and supersaturated with air, and hence in a metastable state with respect to the formation of large air bubbles, but nonetheless survive intact when similar columns of water in typical experimental setups would not.
 
Many details of this highly complex process, for example the physics of hydrogen bonds, the amazing anatomy and ecological evolution of the xylem structure, and the intricate role roots play within the cohesion-tension mechanism, have been discovered or at least better understood since the late 19th century.  Perhaps the most significant development, to which we will return, has been understanding the role of cavitation in the xylem elements, related to the fifth point above. At any rate, even the descriptive language of the CT theory has changed, largely as a result of the introduction of  the ``Ohmic analogy'' of the complete soil-tree-atmosphere continuum in the 1950s, inspired by the physics of electrical circuits, as well as the subsequent widespread use of the phenomenological notion of water potential. In fact, modern statements of the tenets of the theory often use technical terms which would have been unfamiliar to the originators of the theory. 

All scientific theories evolve, but the complexity and multiplicity of the biomechanical processes in trees and the piece-meal advances in the understanding of these disparate processes makes it harder than is often the case in physics to say whose theory it is. But as most historical treatments start in the 19th century, we shall follow suit -- for now.

\subsection{19th century origins}

Although speculation that the ascent of sap might be a purely mechanical process had arisen in the eighteenth century (as we see below), confidence in such a view was by no means universal in the late nineteenth century. Thus we find the the Swiss botanist Simon Schwendener (1829-1919) quoted as saying in 1886:
\begin{quote} 
I absolutely stand by the fact that the vital activity of cells is somehow intervening in sap motion. The lift of water up to heights of 150 to 200 feet and more, is simply impossible without this intervention. And all the endeavours to break through existing barriers by uncertain physical concepts, are not much more than seeking the philosopher's stone.\footnote{See Ehlers and Goss (2003), p. 82, who attribute this quotation  to B\"{o}hm (1893).}
\end{quote}
So it was one of the achievements of the great Polish-German botanist Eduard Strasburger (1844-1912) to provide by 1891 powerful evidence to the effect that living cells were not the causal agents of water ascent inside plants, dealing a significant blow to centuries of speculation that vitalist mechanisms had to be involved.\footnote{See Strasburger (1891, 1893).}  He had further instigated what the distinguished botanist Sir Francis Darwin (son of Charles Darwin) called a ``violent disturbance of our current views''\footnote{Darwin \textit{et al.} (1896), p. 640.} which held that at the time of most active transpiration the xylem conduits contain air, and not water. Strasburger
 demonstrated that sap forms continuous columns from roots to leaves in some xylem conduits, while other such  conduits are filled with air, which do not function as pathways for water. Finally, he was able to show that cause of the the rise of sap 
could not be barometric pressure, nor root pressure, and concluded that no satisfactory physical explanation existed.

The confusion concerning the ascent of sap at this time was well described by Francis Darwin:
\begin{quote}
The ... question [concerning the forces producing ascent] has a curious history, and one that is not particularly creditable to botanists generally. It has been characterized by loose reasoning, vagueness as to physical laws, and a general tendency to avoid the problem, and to scramble round it in a mist of \textit{vis \`{a} tergo} [root pressure], \textit{capillarity, Jamin chains} [a succession of bubbles of air separated by water], \textit{osmosis} and \textit{barometric pressure}.\footnote{Darwin \textit{et al.} (1896), p. 631. Copeland (1902) provides much detail on the range of ideas and experimental evidence being brought to bear on the issue in the 19th century -- and his own skepticism about the validity of the CT theory.}
\end{quote}

Further evidence of the physical, rather than vitalistic, nature of the phenomenon was provided in 1892 by the Austrian botanist Josef Anton B\"{o}hm (1833-1893), a professor at the `Hochschule f\"{u}r Bodenkultur' (University of Agricultural Sciences) in Vienna.\footnote{B\"{o}hm shares his name with the celebrated Austrian sculptor -- anglicized as Joseph Boehm -- whose 1885 statue of Charles Darwin is found in the Natural History Museum in London.} But an even more significant contribution by B\"{o}hm was published in 1893, shortly before his death. He recognised that transpiration was important in drawing water up from the roots, and that water columns form a continuum from the leaves to the soil. He argued that molecular forces between water molecules are more than sufficient to maintain the integrity of the columns despite the tensions involved, and referred to Helmholtz's 1873 work on the cohesive nature of water.\footnote{Helmholtz (1882, p. 830.).} He also speculated (he was not the first) that capillarity is essential for driving the ascent of such columns, but the whereabouts of such capillarity within the plant is tantalisingly unspecific  in the paper. ``It never entered my mind'', B\"{o}hm wrote, ``to think, or to claim, that the [cause of the] rise of sap occurs only in the [xylem] vessels". But this is followed, after a digression, with the rather disappointing statement: ``The sap-bearing wood is ... a very complicated capillary system and capillarity still an unclear area of physics". B\"{o}hm concluded that ``explaining the extent to which concave menisci play a role [in the hypothesised capillarity] is a task for physics."\footnote{B\"{o}hm (1893). The translation from the German of these passages is due to John Pannell (private communication). Useful summaries of Bohm's energetic experimental work on the rise of sap are found in Ehlers and Goss (2003), pp. 82-4.} (This proved to be a prescient remark.) A categorical statement that the principal capillary action takes place in the leaf parenchyma is not to be found in this, nor in B\"{o}hm's earlier papers. But his innovative appeal to the cohesive properties of water made sense, or at least  lessened the counterintuitive nature, of Strasburger's empirical results concerning the existence of continuous columns of water within the xylem.

The next important figures in this story were two Irish scientists. John Joly (1857-1933) was an engineer, physicist and geologist who made many notable contributions to science. These included the development of radiotherapy in the treatment of cancer, the development with Ernest Rutherford of techniques for estimating the age of geological periods based on radioactivity in minerals, and pioneering work in colour photography. His younger collaborator, Henry H. Dixon (1869-1953), was a prominent botanist who like Joly in the late nineties was teaching in Trinity College Dublin.\footnote{Biographical details related to Joly and Dixon are found in Wyse Jackson (1992) and Jones (1992), respectively.} In 1894, Dixon and Joly published the abstract  to a paper on the ascent of sap still under review in the \textit{Philosophical Transactions of the Royal Society of London}\footnote{Dixon and Joly (1894).}, followed in 1895 by the revised paper\footnote{Dixon and Joly (1895).}. It ``diffidently" pointed out that transpiration in the leaves must be sucking the water up all the way from the roots, even in very tall trees. 

Dixon and Joly were struck by the ancient method of evaporative cooling of liquids like wine by placing them in porous pots partly immersed in water. ``A porous pot, in which the pores were so exceedingly fine that the water meniscuses formed in them would be able to support a tension equivalent to many atmospheres pressure, and supplied below with water in a state capable of standing this high tension, would represent (according to our views) the arrangement obtaining in high trees." This classic paper not only referred to a number of careful experiments on the evaporation of water from the leaves of plants under pressure, including ones performed by the authors, but also demonstrated clear appreciation of the major complication in the proposed mechanism: the danger of cavitation in water under tension. This is the breakdown of the tensile strength of such metastable water by way of fracturing, leading to the formation of air gaps, or embolisms. Dixon and Joly showed empirically that water with dissolved air can survive a significant degree of tension without fracturing, and, as we shall see below, they anticipated the rudiments of later work explaining why sap doesn't boil in the xylem conduits. (They gave generous acknowledgement to George Francis FitzGerald for discussions related to the physics of the cohesion theory. FitzGerald, the most gifted Irish physicist of his generation, is famous for his discovery of the so-called FitzGerald-Lorentz contraction phenomenon, one of the central predictions of Einstein's special theory of relativity.) 

Dixon and Joly are often cited today as either the founders of the CT theory, or amongst its founders,  but arguably equal honours should be assigned to Eugen Askenasy (1845-1903), who in the 1890s was Professor of Botany at the University of Heidelberg. Askenasy read the 1894 Dixon-Joly abstract, and, shades of Darwin and Wallace, was spurred to publish a detailed account of the rise of sap that he had independently been developing for some time. Askenasy's insights were in large part the same as those of Dixon and Joly, and were published in 1895 --- before the Irish duo published their main paper --- and in 1896.\footnote{``The most important common feature to both papers was the identification of the cell walls of [leaf] parenchyma cells, whether living or dead, as the sites where surface tensions develop due to the transpiration of water. Both papers emphasized that a moist cell wall is impermeable to air, so that even at negative pressures air cannot be sucked into conducting elements.'' Richter and Cruiziat (2002). It is worth noting that this last point was clearly made in the 1894 abstract by Joly and Dixon. In Brown (2013) it is incorrectly stated that the relevant pair of Askenasy's papers appeared in late 1894 and 1895.} According to a 1902 historical review published by the American botanist Edwin Bingham Copeland, it was Askenasy's clear formulation of the theory that was originally responsible for its widespread acceptance.\footnote{Copeland (1902). See Askenasy (1895, 1896).}

\subsection{Priority issues}

In one respect it seems that Askenasy let wishful thinking get the better of him. Having seen that the 1894 Dixon-Joly abstract was somewhat tentative about the exact nature of the lifting force, Askenasy concluded that these authors had subsequently adopted his own 1895 account of the role of capillarity in certain cell walls of the transpiring leaves. Thus, in an address to the Liverpool Meeting of the British Association in September of 1896 which featured a discussion of the rise of sap, Joly took up what he called the ``distasteful" matter of priority, which by that time had become a source of tension between Dublin and Heidelberg. Joly made a point of clarifying that the role of capillarity in cell walls of the leaves was independently proposed in the original Dixon-Joly manuscript. He speculated that Askenasy may have been misled by the fact that their 1895 publication was described as a revised version of the original manuscript, when in fact the revisions involved no new material, only deletions.

Apart from capillarity, the role of osmotic forces in the leaves in accounting for the ascent of sap (as opposed to the maintenance of rigidity of the leaves despite the internal tensions due to transpiration) is today largely discredited, but for a considerable period it was also taken seriously by our protagonists.\footnote{For details, see the discussion in Darwin \textit{et al.} (1896); see also Pickard (1981), p. 220.} Thus Pickard was led to remark in 1981:
\begin{quote}
The question of priority is of far less importance to the modern reader than the fact ... that, before the theory could fairly be said to have passed out of its perinatal stage, Askenasy, Dixon and Joly had all commenced a vascillation over the origin of suction force in the leaf: whether the agency which upheld the cohering column of sap was capillary imbibition by the walls of the evaporating cells or whether it was osmotic action by the cells adjacent to the upper reaches of the water conduits. Vestiges of this doubt have persisted to the present day.\footnote{Pickard (1981) p. 185.}
\end{quote}

Be that as it may, Joly asserted in 1896 that he and Dixon had ``conceived of our theory quite originally, being
unaware of any suggestion even remotely resembling our explanation of the cause of the ascent of sap in high trees.'' This claim deserves a little scrutiny. 

Dixon and Joly cite B\"{o}hm's 1893 work in their original paper (and indeed B\"{o}hm's 1983 paper was apparently discussed when the two Irish collaborators visited Strasburger in that year.\footnote{See Richter and Cruiziat (2002).}) They find ``perfect agreement'' between their own experiments, showing that significant transpiration can take place in branches even when external pressures of 2 to 3 atmospheres are applied, and B\"{o}hm's success (after many attempts) in arranging for a transpiring branch to lift a column of 90cm of mercury. So water continues to reach the leaves whether the air-water interfaces in the branches are being pushed, or pulled, back into the branch, and Dixon and Joly conclude that the forces associated with transpiration are sufficient to lift tensile columns of raw sap into leaves of tall trees. In this they do not appear to differ qualitatively from B\"{o}hm; indeed all parties agree that cohesion and capillarity are key elements in the story. 

Considerations such as these have understandably led some commentators to decry the drowning out of B\"{o}hm's name in the debate about priority that occurred in the years following his death.\footnote{\textit{Ibid}. Copeland (1902), p. 191, makes this droll remark: ``Ten years ago B\"{o}hm alone imagined (publicly) that capillarity could play the leading role in the ascent of sap. It had been tried and found wanting. Then it was named cohesion and sprang at once into popular favor.''}
The omission is evident in, for example, the contribution by Francis Darwin to the aforementioned 1896 Liverpool meeting. In his detailed review of theories of the rise of sap (which contains very useful historical comments) Darwin stated:
\begin{quote}
In 1894 Messrs. Dixon and Joly first enunciated an entirely new theory, depending upon the quality which water possesses of resisting tensile stress. To most botanists the existence of this quality is a new idea. To believe that columns of water should hang in the tracheals like solid bodies, and should, like them, transmit downwards the pull exerted on them at their upper ends by the transpiring leaves, is to some of us equivalent to believing in ropes of sand. Meanwhile Askenasy had independently hit on a similar theory which was published after the appearance of Messrs. Dixon and Joly's research.\footnote{Darwin \textit{et al.} p. 635.}\end{quote}
In the 1896 meeting Joly himself pointedly referred to a remark of Askenasy:  ``As I have remarked in my first paper, Dixon and Joly are the first who have clearly recognized and formulated the significance of the cohesion of water in the ascent of sap.''\footnote{Darwin \textit{et al.} (1896), p. 648-649.} In more recent times, however, the seminal contributions of Josef B\"{o}hm to the birth of the CT theory are often cited; some commentators describe him as the formulator of the theory.\footnote{See, for example, Steudle (2001), p. 854.} 

But we should not lose sight of what Dixon and Joly did that B\"{o}hm did not.
It is clear from their paper that Dixon and Joly regard the crux of their novel contribution as the clarification of the role leaves play in the ascent of sap.  They stress that the ``all-sufficient cause of the elevation of sap'' is suction by the leaves, not ``by establishing differences of gas-pressure'', as some botanists had already claimed (though not presumably B\"{o}hm), but ``by exerting a simple tensile stress on the liquid in the conduits''.\footnote{An account of the process of elimination used by Dixon and Joly in the months before arriving at their final theory in 1894 is found in Dixon  (1914), Ch. IV. A series of existing theories were analysed and rejected, including gravitational and electrical ones; a combination of Quinke's theory of thin films of water being drawn up the xylem walls with the Unger-Sachs imbibition theory was also entertained and discarded.}  In his contribution to the 1896 Liverpool discussion, Joly stressed the difference between the roles of capillarity and evaporation in the new theory.  Suppose a steady state is reached where water loss in leaves due to transpiration is replaced by flow through the xylem. In that case,  Joly points out, capillarity plays no ``active'' part:
\begin{quote}
The attraction upon the water as much acts to retard evaporation as to elevate the sap. It is a purely static and undirected stress, and does no work .... [T]he only kinetic action effective in actually lifting the water -- in fact the only directed action -- is that derived from the directed energy of those water molecules which escape from the evaporating surface or meniscuses ; their loss being made good by molecular attractions and diffusion forces acting within the liquid. These are the actual lifting forces; the work done in lifting the water against gravity and viscous resistance being but small compared to the normal work of evaporation. The energy expended by the liquid is finally restored by the inflow of heat at the evaporating surface.\footnote{Joly  is here (Darwin \textit{et al.} (1896) p. 652) referring to a model involving a porous pot connected to a supply of water, but the argument is supposed to illuminate what happens in trees.}
\end{quote}
For these and other reasons (in particular, the role of osmotic pressure in maintaining the rigidity of leaf cells during transpiration), Joly disassociated his and Dixon's position from Askenasy's claim that the imbibition/capillarity forces of the cell-walls in the leaves are alone ``the so long vainly sought for source of the suction-force concerned in the ascent of sap in the plant''. The reader may wish to compare Joly's reasoning with a more recent analysis due to Pickard:
\begin{quote}
Evaporation on the other hand is not the ultimate force either. It merely abstracts molecules from an interface thereby altering its shape. And it is adsorptive forces at the meniscal perimeter and surface tension forces over the interface itself which, in striving to restore the interface to its equilibrium configuration, exert the tension. Not surprisingly, these distinctions are difficult to grasp (and no less difficult to explain!)! and the roles of the leaf cell as an osmometer and of osmosis as that agency which opposes the loss of both intracellular water and leaf turgor in the face of evaporation have not been as well appreciated as might have been hoped.\footnote{Pickard (1981) pp. 220-221.}
\end{quote} If there is any disagreement here between Joly and Pickard, it may have more to do with emphasis than substance. When several processes are involved -- evaporation, capillarity, molecular attractions and diffusional forces -- the distinctions between which are not categorical, it may be more important to recognise their joint actions than to specify the nature of ``the ultimate force''. There can be little doubt, at any rate, that Dixon and Joly, aided possibly by the discussions with FitzGerald, had a much better grasp of the physics of capillarity and surface tension than B\"{o}hm, of how the phenomenon of evaporation in the leaves gives rise to a lifting force and what its magnitude is, and of the cooling effect and overall thermodynamics of evaporation. 

Joly himself emphasised that no previous theories had adequately addressed the issue of the stability of tensile threads of water containing dissolved air and in contact with wetted wood, and that no one had \textit{quantitatively} shown how the suction in transpiring leaves in high trees is capable of lifting such threads of sap. He also remarked on the asymmetry between leaves and roots. Evaporation in leaves takes place because the water vapour pressure in the air is less than that in the substomatal cavities. Condensation of water vapour from the soil onto the roots takes place because here the opposite is true. So `` ... we can regard the whole phenomenon of the ascent of sap to depend broadly upon a difference of vapour pressure in the atmosphere surrounding the roots and in the atmosphere surrounding the leaves.''\footnote{\textit{Op. cit.} p. 659.} Here, Joly was unaware that liquid phase uptake in roots is normally much greater than vapour absorption: the hydraulic continuum extends into the soil pore space.\footnote{I owe clarification on this point to John Sperry.} It should not be overlooked too that the possibility of vitalist extrusion mechanisms in the leaves supposedly acting in parallel with transpiration was suggested by Joly in 1896 and Dixon in 1914 -- a possibility which today is disregarded.

In his 1914 monograph, Dixon had this to say about B\"{o}hm's final theory of the ascent of sap:
\begin{quote}
Owing to his contradictory expressions and obscurity in description it will always remain impossible clearly to understand what his hypothesis was. He assigned a part to the capillary forces of the tracheae and to atmospheric pressure. The latter he conceived as driving the water from the tracheae into the leaf-cells, and also he saw no difficulty in the height of the water columns owing to the cohesion of water; but without doubt he was quite astray as the conditions under which cohesion could act.\footnote{Dixon (1914), p. 29. It is noteworthy that the five B\"{o}hm papers cited by Dixon do not include the 1893 paper.}
\end{quote}
Whether this judgement of B\"{o}hm's work is excessively harsh is a matter for future scholars -- as is the influence B\"{o}hm  may have had on Askenasy, a topic left untouched here.

\section{When is a theory born?}

\subsection{The Newton of plant physiology}

In 1999, in a letter to the journal \textit{Trends in Plant Science}, Franz Floto claimed that ``all the elements'' of the theory associated with the names of B\"{o}hm, Dixon, Joly, and Askenasy were first described in 1727 by Stephen Hales in his book \textit{Vegetable Staticks}. Occasionally, within modern historical outlines of the theory of the rise of sap, mention is indeed made of the (literally) ground-breaking experiments of this remarkable English clergyman-scientist, ``the Newton of plant physiology''\footnote{Harr\'{e} (1970) p. 65; this work contains a short biography of Hales in Chapter 7; amongst his many contributions to science was the discovery of carbon dioxide in air.}, related to the movement of water in plants --- an extension of his detailed research into the circulation of blood in animals. But Floto goes beyond what seems to be the standard view of Hales' contributions, and describes him as the forgotten father of the CT theory, overlooked particularly in the 19th century.\footnote{Floto (1999).}

The matter deserves careful historical study, but a few tentative remarks may be in order. Hales' elegant and painstaking experiments, which included the first use of trace-elements in the history of botany, and one of the first uses of sampling techniques, do seem to have been less widely known in 1830 than 1730.\footnote{Some of the following comments are based on the helpful analysis of these experiments found in Harr\'{e} (1970), chapter 8.} Amongst the results were demonstrations that the principal driving force of the ascent of sap is transpiration from leaves rather than a positive force from the roots; that it is indeed water (if somewhat impure) that is transpired from leaves; that rain, not dew, provides the soil (in typical English conditions) with enough moisture for trees to survive; that trace elements in sap do not impregnate fruit; that a small amount of air in the stem conduits can significantly impede the flow of sap, that sap does not circulate in the tree as blood does in animals, and that some part of the nourishment of a plant is probably provided by air entering the leaves. Furthermore, Hales showed by way of construction of an ``Aqueo-mercurial gauge'' of his own design that evergreens have less power to imbibe water than deciduous plants, and that generally the absence of sunlight, as well as that of leaves, reduces the rate and degree of suction of water. It was clear to Hales that the rise of sap in plants was a mechanical, non-vitalistic process consistent nonetheless with the absence of an ``engine, which, by its alternate dilatations and contractions'' would pump water as the heart does in animals; and `` ... yet has nature wonderfully contrived other means, most powerfully to raise and keep in motion the sap.''\footnote{Hales (1727), p. 43.} On the role of the importance of ``perspiration'' (transpiration) from leaves, Hale explicitly theorised within the context of Newton's discussion in the \textit{Opticks} of the capillary rise of liquids like mercury and water:

\begin{quote}
And by the same [Newtonian] principle it is, that we see in the preceding Experiments plants imbibe moisture so vigorously up their fine capillary vessels; which moisture, as it is carried off in perspiration, (by the action of warmth), thereby gives the sap vessels liberty to be almost continually attracting fresh supplies, which they could not do, if they were fully saturate with moisture: For without perspiration the sap must necessarily stagnate, not withstanding the sap vessels are so curiously adapted by their exceeding fineness, to raise the sap to great heights, in reciprocal proportion to their very minute diameters.\footnote{p. 56. It seems that Hales himself may not have been the first to connect capillarity with the rise of sap; Dixon (1914), p. 27, refers to the 1723 views of Christian Wolff who ``believed that the forces involved were the expansion of air and capillarity''.}
\end{quote}

Now were Newton to have had an understanding of the microscopic forces underpinning capillarity, Hales might have also have better appreciated the role of cohesion of water itself in the whole story. It is perhaps the lack of emphasis on this feature, and the lack of explicit clarification that water forms \textit{continuous} tensile threads from roots to leaves, that first militates against the view that Hales, despite all his remarkable insights, was the progenitor of the CT theory.  
He did not and could not have appreciated the role of hydrogen bonds between water molecules and between water and wettable surfaces, but the same can be said of B\"{o}hm, Dixon, Joly, and Askenasy --- the first intimations of the nature of hydrogen bonds having occurred in the 1920s\footnote{See Rowlinson (2002), section 6.3}. These scientists appealed to the extraordinary cohesion and adhesion of water as empirical facts -- ones which were in the air in the late 19th century. 

As it happens, the discovery that water and other liquids could exist in states of tension predates Newton's publications. In 1661,  Christian Huygens realised that water could exist metastably at a pressure lower than vapour pressure. Aware of Huygens' discovery, Robert Boyle in 1663 showed that mercury could stick to the glass in a barometer and support a tension of two atmospheres. These phenomena were known to Thomas Young and Pierre-Simon Laplace, who independently at the start of the 19th century were to develop theories of surface tension and cohesion of liquids.\footnote{For an account of their work, particularly related to capillarity, see Rowlinson (2002).} M. F. Donny's 1846 experiments and observations on the tensile nature of liquids were similar to those of Huygens and Boyle, but it seems Donny was unaware of these precedents. In fact the history of the science of liquids under negative pressure appears to be one of repeated rediscovery.\footnote{Details of the early work in the field can be found in Kell (1983).} In 1882, Osborne Reynolds (of ``Reynolds number'' fame) published results showing negative pressure in mercury, in what he originally thought was a new phenomenon, but by the time of publication he had been made aware of Donny's work. Reynolds was apparently unaware of Marcelin Bethelot's 1850 investigations of negative pressures using a different technique, and one which became a powerful research tool in the 20th century after having been overlooked for many years.\footnote{See Trevena (1978).}

But it was B\"{o}hm in 1893, not Hales in 1727, who had first seen the connection with the cohesive nature of liquid water and the rise of sap in trees. 
Hales unfortunately entertained the possibility, as Francis Darwin noted in 1896, that water may travel as vapour from roots to leaves, not in the liquid state, at least in vines.\footnote{Hales (1727) p. 9; Darwin \textit{et al.} (1896), p. 630.} 

One might also be tempted to say that Hales put capillarity in the wrong place. Following Askenasy, and Dixon and Joly, the relevant locus of capillarity is supposedly in the  leaves. It is common in accounts of the CT theory to find the point made at the start of section 2 above, that typical diameters of xylem conduits in trees are incompatible with capillarity-driven heights of more than a few meters.\footnote{See in particular Steudle (1995), who suggests that non-botanists typically make the mistake of attributing sap rise to, if anything, capillary action in the xylem.} But there is more to it than meets the eye, as we shall see later.

\subsection{A useful distinction}

What arguably is commendable in the above passage from \textit{Vegetable Staticks}, and what seems to be widely missing  from modern accounts of the rise of sap -- at least from the perspective on non-plant scientists -- is the distinction being made between two issues: \textit{how sap gets to leaves, and how the flow is accelerated under transpiration}. It is one thing to account for the relatively high rate of flow of sap in the stems of plants in daylight hours; it is another to show how sap happens to be in the leaves in the first place. For Hales, the two issues have very different origins, as can be seen in the above passage and the following, again from \textit{Vegetable Staticks}:
\begin{quote}
... in vegetables we can discover no other cause of the sap's motion, but the strong attraction of the capillary sap vessels, assisted by the brisk undulations and vibrations, caused by the sun's warmth, whereby the sap is carried up to the top of the tallest trees, and is there perspired off thro' the leaves: But when the surface of the tree is greatly diminished by the loss of its leaves, then also the perspiration and motion of the sap is proportionally diminished, as is plain from many of the foregoing Experiments: So that the ascending velocity of the sap is principally accelerated by the plentiful perspiration of the leaves, thereby making room for the the fine capillary vessels to exert their vastly attracting power, which perspiration is effected by the brisk rarifying vibrations of warmth ... \footnote{Hales (1727), pp. 77-8. It is hard to reconcile this passage with the claim by Beerling and Frank (2010) that Hales, despite being aware of the role of ``perspiration'' in the ascent of sap, did not discover that plants transpire water from leaves.}
\end{quote}
Hales' distinction between the initial motion of sap and its subsequent acceleration through transpiration is particularly relevant in the well-documented case of certain vines which reach considerable heights using trees as vertical supports, and which become air-filled in winter.\footnote{Studies include that of the common northern grapevine (\textit{Vitis labrusca}) which can sometimes attain heights of 17 to 18 meters in wooded areas in Massachusetts; see Scholander \textit{et al.} (1955).}  The rise of sap in the spring occurs \textit{before the appearance of leaves}, and is commonly attributed to positive root pressure.\footnote{It is noteworthy that Hales (1727), Ch. III, performed the first published experiments designed to measure root pressure, in grapevines (\textit{vitis vinifera}). Another, perhaps complementary, positive-pressure mechanism in tall vines, based on temperature-associated osmotic water uptake from rehydrated cells of the bark, was suggested in Pickard (1989).} 

As mentioned earlier, root pressure is known to be insufficient to account for the existence of sap in the leaves of tall trees generally.\footnote{See Pickard (1989), p. 264, and Tyree \textit{et al.} (1999).} But the role of capillarity in the xylem conduits does seem to be a significant element in the overall story, as was stressed by Park S. Nobel in 2005:
\begin{quote}
The numerous interstices in the cell walls of xylem vessels form a mesh-work of many small, tortuous capillaries, which can lead to an extensive capillary rise of water in a tree. A representative ``radius'' for these channels in the cell wall might be 5 nm ... [which] could support a water column of 3 km -- far in excess of the needs of any plant. The cell wall would thus act as a very effective wick for water rise in its numerous small interstices, although the actual rate of such movement up a tree is generally too low to replace water lost by transpiration. 

Because of the appreciable water-wall attraction that can develop both at the top of the xylem vessel and in the numerous interstices of its cell walls, water already present in the lumen of a xylem vessel can be sustained or supported at great heights. The upward force, transmitted to the rest of the solution in the xylem vessel by water-water cohesion, overcomes the gravitational pull downwards. \textit{The key to sustaining water already present in the xylem vessel against the pull of gravity is the very potent attractive interaction (adhesion) between water and cell wall surfaces in the vessel.} [emphasis HRB]\footnote{Nobel (2005).} 
\end{quote}

The last claim in this passage, that capillarity in the interfibrillar spaces within xylem walls sustains water columns against gravity,\footnote{A similar statement is found in Pickard (1981): ``... the less than 100 nm interstices of the cell wall will support without difficulty the 100 m long water columns of the largest redwood tracheary elements.'' In the same vein (no pun intended), Steudle (2001) writes: ``Xylem walls consist of a porous net of wettable polymers (cellulose, lignin, hemicelluloses, etc). Pores (interfibrillar spaces) are of an order of 10 nm, which corresponds to 30 MPa of capillary pressure. Hence the porous hydrophilic matrix will be imbibed with water like a sponge.'' In fact, stress on the role of adhesion between water and xylem walls in sustaining hanging threads of water is already found in Dixon (1914), who cites experiments performed with Joly (discussed below) as showing that such adhesion is stronger than that between water and glass. ``This is quite to be expected, when we take into account the manner in which water permeates the substance of the walls of the tracheae when brought into contact with them.'' p. 87.} in itself represents a partial vindication of Hales' insights --- though the precise nature and details of the capillarity/adhesion are something Hales could not anticipate. The first paragraph in the passage goes even further in the Halesian direction and seems to suggest that such capillarity can account for the ascent of sap, though not of course at the flow rates associated with periods of transpiration. However, to appeal to the wick-effect in order to account for hydration of the xylem walls is one thing; the claim that it accounts for filling of the conduits themselves is another.\footnote{See Pickard (1989), p. 268.} Presumably the putative role of capillarity-driven sap rise will depend on the type and height of the plant; its role in the case of tall trees seems to find little support in the literature.\footnote{This capillarity view has a certain amount in common with the ``imbibition'' theory of sap rise that was prominent prior to the 1880s and largely associated with the name of the great German botanist Julius von Sachs, though this theory held that the flow of sap takes place within the walls of xylem conduits; for details see Copeland (1902), pp. 179-181, and Dixon (1914) Ch IV. It is noteworthy that in 1896, the happily-named Vines was measuring significant tension within leafless branches, and speculated that Sach's theory ``includes some at least of the essential elements of a complete explanation''; see Vines (1896) and his contribution to Darwin \textit{et al.} (1896) pp. 644-647.}

But if the question is how sap finds itself \textit{from the beginning} within the entire vascular structure of the tree or vine, one crucial issue has yet to be addressed. It behoves us therefore to turn, somewhat counterintuitively, to the question as to why sap does not disastrously boil, and its columns break, in trees. 

\section{Avoiding and recovering from cavitation}

\subsection{Why so little cavitation?}

As we have seen, empirical evidence of the cohesive nature of water was building up in the 19th century -- largely unbeknownst to botanists.\footnote{For remarks on the farsighted 1886 work of N\"{a}geli, see Copeland (1902).} But experimental tests to determine precisely how much tension water can sustain before fracturing have proved to be notoriously unreliable, giving rise to very erratic results.\footnote{See in  particular the survey of results in Greenridge (1957).} This is essentially due to the fact that for anything other than extreme tensions, fracture occurs not because catastrophic breakage of hydrogen bonds is taking place, or because of the spontaneous formation of bubbles containing water vapour, or previously dissolved air,  but rather because of the prior existence of tiny bubbles of air or the liquid's vapour --- ordinary boiling, in short.\footnote{``... the distinction between cavitation (normally, fracture at room temperature and negative pressure) and boiling (normally, fracture at elevated temperature and atmospheric pressure) is easily blurred since both are aspects of the single phenomenon of nucleus growth.'' Pickard (1981), p. 200.}  Since the seeding  of such bubbles varies with the equipment being used, it may not be an exaggeration to say that ``what is measured is not the tensile strength of water, but the size to which bubbles had been reduced prior to the experiment".\footnote{Tyree and Zimmerman (2002), p. 62.}

It was known before the articulation of the CT theory both that the boiling point of water depends on the ambient pressure, and that ordinary boiling requires pre-existing nuclei in the liquid, i.e. tiny pockets of gas in minuscule crevices within the walls of the container.\footnote{See Donny (1846); a fascinating treatment of the history of the theory of boiling is found in Chang (2007).} Unless the situation in xylem conduits is very different from water put under similar tension in ordinary containers, it should boil, leading to cavitation ---the rupture of the threads running from the roots to the leaves. Why is this expected behaviour suppressed inside trees? In a seminal paper published in 1971, J. J. Oertli wrote that ``no adequate explanation of the cohesion [CT] theory can be given without consideration of the theory of bubble formation.'' 

It is now accepted, on the basis of theory and experiment, that for the tensions likely to occur in tree xylem, spontaneous bubble formation (homogeneous nucleation), even where sap is saturated with air, is highly unlikely.\footnote{See Oertli (1971) and Pickard (1981), section IV.} The cause of cavitation would have to be the presence of air nuclei within the interstices of cell walls within the xylem conduits (heterogeneous nucleation). It proves to be very difficult to create similar tensions within experimental conditions using, for example, glass or metal tubes, even when polished, because heterogeneous nucleation is all but unavoidable within microfissures of the walls of the tubes. \textit{It would be very difficult to create a tall artificial tree.}\footnote{This is of course a wild understatement. Besides its hydraulic prowess, a tree of any height produces all the complex molecules needed for its physical structure and metabolism -- including the hormones (auxins) that tell the stem to grow upwards and the roots to grow downwards -- from only sunlight, carbon dioxide, water, nitrogen and a few trace elements. All of this is far beyond human manufacturing capabilities; see Suzuki and Grady (2004), Introduction.} What is Nature's trick? It is not that the walls of xylem elements are perfectly smooth, as was seen in the last section; there are plenty of tiny nooks and crannies for air seeds to hide. 
The trick is \textit{not to allow the introduction of air in the first place}, and this happens because xylem cells are born wet, and ultrafiltration in root membranes stops air pockets from entering the xylem.
\begin{quotation}
It is the presence or absence of ... [gas] nuclei that explains the difference between the plant system and the vacuum pump in which a Torricellian vacuum forms. The conductive tissue has, from the time of its formation, been filled with aqueous solution ... [except in the case of freezing]. Cell division, formation of primary and secondary cell walls, cell elongation, disappearance of cross-walls in long vessels, all occur in an aqueous medium.\footnote{Oertli (1971), p. 207. The point seems to be widely accepted. ``The conditions for heterogeneous nucleation at preexistent sites appear to be equally hard to fulfill: the tracheary element is presumably water filled from earliest differentiation with the result that its putatively hydrophilic boundaries should be thoroughly wet and unlikely to stabilize nuclei ... ''. Pickard (1981) p. 205; see also p. 195.  ``Xylem conduits are water filled from inception and contain no entrapped air bubbles that could nucleate cavitation.'' Tyree and Sperry (1989), p. 20.} 
\end{quotation}

By the time xylem conduits function in a vascular sense, their cells have died and the xylem has become lignified and stiffened, so as to withstand the tensions involved in sap ascent. But the growth of xylem cells, like all plant cells, works mostly by increasing water content.\footnote{Hales himself seems to have had at least a  rudimentary understanding of this point: ``But whether it be by an horizontal or longitudinal shooting, we may observe that nature has taken great care to keep the parts between the bark and wood always very supple with slimy moisture, from which ductile matter the woody fibres, vesicles and buds are formed. ...
We see here too that the growth of shoots, leaves and fruit, consists in the extension of every part; for the effecting of which, nature has provided innumerable little vesicles, which being replete with dilating moisture, it does thereby powerfully extend, and draw out every ductile part.'' Hales (1727), p. 194.} The fact that the xylem element is water-filled from earliest cell differentiation may be so obvious to plant scientists that its wider importance in the story of the rise of sap is seldom stressed --- wider, that is, than detailed discussions of the cavitation issue. But to a physicist, say, wishing to understand how water gets to the foliage of tall trees, the fact surely is of considerable relevance, even it is far from the whole story. 
As Kramer and Boyer succinctly put the point: ``Water moves to the tops of plants as they grow and transpiration merely increases the quantity and speed of movement.''\footnote{Kramer and Boyer (1995), p. 203. It should be noted that this remark occurs in the context of a discussion of how important transpiration is for the growth of a plant (see foonote 19 above).} Another of the somewhat rare instances of emphasis of this point is found in the 2011 review paper by John Sperry: 
\begin{quote}
Importantly, as the [xylem] conduits mature and lose their protoplasts to become functional, they start out full of fluid. If not, their lumens would be unable to generate sufficient capillary suction to fill themselves with water in most conditions. ... Although capillary forces are responsible for driving the transpiration stream, capillary \textit{rise} is not involved. ... [The wick-like nature of a xylem conduit] cannot function unless it is primed by being water-filled. 
\end{quote}

Admittedly, the wick-like feature of xylem conduits, with their porous nature and concomitant  high capillary forces, may be such that ``no space will be left for air-seeds of sufficient size to cause embolism'', as Stuedle has remarked.\footnote{Steudle (2001), p. 854.} But if xylem conduits are wet from their inception and any air seeds are filtered out in the root system before entering the xylem stream, this point must be of questionable importance. The pertinent point is that the nature of cell growth in plants and the filtration process in the root membrane jointly account for water reaching the extremities of trees, and that the biochemical and biophysical processes producing the lifting forces (which include osmosis as well as capillarity) have nothing to do with transpiration.\footnote{For a review of the mechanism of cell growth in plants, and the role of osmosis therein, see Boyer (1985), pp. 492-500, and Boyer \textit{et al}. (1985). It has been pointed out to the author by John Sperry that in in cell growth, as opposed to transpiration, the air-water meniscus is stressed from below, but in both cases the pressure drop is most proximally generated by the air-water meniscus by way of capillary action.} It is true that solutes needed for cell growth are provided by axial flow from the phloem, the vascular system in the inner bark that allows for downward flow from the leaves of sugar-rich sap.\footnote{Unlike the rise of xylem sap resulting from transpiration, the downward transport of sap in the phloem involves a positive pressure osmotic flow mechanism.} But the growth process itself induces water movement even (and preferentially) when transpiration ceases, for example when the plant is surrounded by saturated air in the dark, and normally part of that water is extracted from the soil.\footnote{See Boyer (1988).} Growth, as well as transpiration, pulls water up the xylem.

\subsection{Dixon and Joly again}

It is remarkable that in their original 1895 paper, our Irish heroes had already shown significant insight into the special nature of xylem walls in the context of cavitation. 

\begin{quotation}
Professor FITZGERALD, and, in a written communication, Professor WORTHINGTON, suggested that solid objects in air-containing water might, when this was under tension, collect air upon their surfaces and so bring about rupture. Even if this was so for hard solids, such as glass -- and the results of our comparative experiments did not point to it -- it appeared most improbable that it could apply to a substance so permeated and little differentiated from the water as woody cell-wall.\footnote{Dixon and Joly (1895).}
\end{quotation}
Indeed, Dixon and Joly describe experiments they performed showing that thoroughly pre-wetted wood chips do not act as the sites of rupture when immersed in water which is subjected to tension (by cooling).\footnote{These results were reiterated in Dixon (1914), p. 87.}

Dixon and Joly do not at this point explicitly address the question as to what process is responsible for pre-wetting the xylem walls in the case of plants. In a footnote, however, they justify the notion that sap may contain dissolved air by appealing somewhat obscurely to root pressure. Later in their paper, they note that the time of year in which the tree has its highest water content is spring, when buds start to unfold, a fact which again is connected with root pressure:

\begin{quotation}
The fact that gas bubbles do occur in the region of the wood which transmits the
ascending current indicates an important function of root-pressure, i.e., to dissolve up and clear out the gaseous contents of such conduits as are occupied with bubbles. In the autumn time the gaseous contents of the tree increase in amount, and in the spring the young buds are provided with sap, and at the same time continuity in the watery contents of many of the conduits is re-established, \textit{so that at the beginning of active transpiration there may be a water column of as large a section as possible existing in the conducting wood.}

This view, namely, that the functioning conduits must necessarily have their lumina swamped with water, is supported by another observation of STRASBURGER's, of a different kind. He found that, in order to make air-dried wood capable of conducting water, it was necessary, in addition to thoroughly soaking the walls, to inject the lumina also with water, which would, of course, have the effect of establishing continuous water-columns in the wood. (emphasis HRB)
\end{quotation}

What is noteworthy here is that Dixon and Joly are, like Hales before them, suggesting distinct mechanisms behind the spring \textit{filling} of the xylem conduits so as to establish the continuous columns from roots to leaves, and the summer \textit{flow}. The joint claim is that the filling mechanism is at least in part root pressure, and that filling is a condition for the subsequent ascent. 

It is interesting that in relation to vines and some trees, Dixon and Joly's original idea of root pressure clearing out gas from xylem conduits in spring has been borne out. It might be expected here that for vines, conduit walls at the end of winter have enough air seeding for catastrophic cavitations; yet water conduction must start to take place in existing, refilled conduits -- the growth of new ones only comes about after the expansion of leaves. The evidence does seem to show that root pressure in spring removes cavitation nuclei by the separate mechanisms of dissolving gases and expelling them from the vine.\footnote{See Tyree and Sperry (1989). Apart from vines, root pressure can act so as to recover embolized xylem conduits in herbs, shrubs and small trees; it has never been recorded in gymnosperms or in most forest trees. See Tyree \textit{et al.} (1999).} At any rate,  it is surely clear that the capillary/cohesion theory of the rise of sap that Dixon and Joly proposed in 1895 addressed itself not to the question of how sap gets to leaves \textit{simpliciter}, but to that of how the transpiration stream works. At the end of his 1914 monograph, Dixon wrote:
\begin{quote}
Therefore when root pressure is not acting and when the leaves are transpiring, the cohesion of their  sap explains fully the transmission of the tension downwards, and consequently explains the rise of the sap.\footnote{Dixon (1914), p. 210. But there is a certain tension in Dixon's narrative. On p. 89 we read: ``When allowance is made for the resistance opposed by the conducting tracts to the motion of water in them, we must conclude, that the supply of water raised by these forces [root-pressure and atmospheric pressure] to a height of 10 metres above the roots, must be exceedingly small. It follows that the water in the tracheae above this level is at all times in tension, and, in times of vigorous transpiration, whenever the loss cannot be made good by the lifting pressure of the atmosphere, the water in the tracheae of leaves, at lower levels also, is in a tensile state.'' This picture of course stands in need of revision at least in the light of the annual drying out of tall vines. But on p. 95, following a discussion of the cavitation effects due to bubble formation in the xylem, Dixon concedes that ``... the periodic flooding of the tracheae with water forced upwards by root-pressure will bring the bubbles into solution and will re-establish the conditions for tension throughout the water-tracts''. Root pressure is thus not an insignificant factor after all. And in his final Summary, Dixon states (p. 210) that the  ``configuration, physical properties, and structure of the wood compel us to admit that the water in the conducting tracts, when not acted upon by a \textit{vis a tergo}, must pass into a state of tension. This state is necessitated by the physical properties of water when contained in a completely wetted, rigid and permeable substance which is divided into compartments.'' The next sentence is that cited in the main text above. 

Dixon still does not explicitly explain how the xylem is wetted, but for heights greater than 10m, root pressure seems to be the only candidate in his 1914 account. And as noted earlier,  in his account, transpiration is not the only lifting mechanism in the leaves; a vitalist notion of secretory actions taking place in cell walls is also upheld, which dominates transpiration when the surrounding air is saturated.}
\end{quote}

\subsection{Why so much cavitation?}

By 1981, the understanding of the physics of cavitation inside the xylem had developed to the extent that the mystery was now why it appears to happen as much as it does!\footnote{See Pickard (1981), pp. 205-208.} When a sap column in the xylem snaps under tension, it creates an acoustic emission of energy. Starting in the 1960s, J. A. Milburn and others detected audio-range (low frequency) emissions in a wide variety of species. Later, Melvin T. Tyree and collaborators  developed a more powerful technique using ultrasonic frequencies, and John S. Sperry introduced a method of measuring embolism by way of reduction of hydraulic conductivity. The combined evidence is that is that cavitation is extensive in Nature after all.\footnote{See Tyree and Sperry (1989).}

The two principal conditions responsible for cavitation are understood to be dehydration and (for some species) freezing. Dehydration leads to increased tension inside the xylem, and eventually an air bubble can be pulled from an air-filled conduit into an adjacent water-filled one. The bubble then expands to fill the conduit and water transport fails. Otto Renner first proposed this process of ``air-seeding'' in 1915, and it was later advanced by Martin Zimmermann in 1983; we return to the mechanism shortly. In the case of freezing, gases previously dissolved in the sap escape because of their low solubility in ice. As the temperature rises, depending on the conditions this gas can either either dissolve back into the xylem sap or expand to obstruct the xylem conduit, in the latter case again resulting in an embolized dysfunctional conduit. In (ring-porous) species such as oak and ash, widespread impairment of xylem conduits may occur after a single freeze-thaw event; in other (diffuse-porous) species like maple and beech, loss of hydraulic conductivity develops gradually over the winter.\footnote{In one study of sugar maple saplings, 80\% loss of conductivity in upper twigs, and 60\% in trunk xylem, was reported. See Tyree and Sperry (1989).} In conifers, such as pines, redwoods and spruces, it seems that freezing produces little cavitation, and in walnut trees the embolism rate actually goes down in winter, just to show that nothing can be taken for granted in the plant world.\footnote{For further details on the nature of cavitation, see Tyree and Zimmermann (2002), section 4.1.}

A remarkable feature of xylem structure is that rising sap has ways of getting round obstructed conduits. (Pits in the walls between adjacent conduits allow for flow between them. Except in the drastic conditions that give rise to the air-seeding mentioned above, the pits are capable of preventing passage of an air-water meniscus in cases where one conduit becomes air-filled.) The most dramatic evidence of this is the almost negligible effect that even deep transverse saw cuts across tree trunks and branches can have on transpiration, as Stephen Hales, and many following him, have noted.\footnote{Hales (1723) and others, most famously R. D. Preston in 1952, have investigated the effects of pairs of transverse cuts separated longitudinally but coming from opposite sides and overlapping. Little effect on transpiration takes place if the distance between such cuts is greater than a critical distance, which is roughly double the typical length of the xylem conduits in the given tree. Single transverse cuts severing up to 90\% of the cross-sectional area of the xylem likewise have little effect on transpiration in some cases, which is consistent with the belief that the stem of a woody plant has an hydraulic resistance that is small compared to other parts of the water pathway in the plant. For further discussion, see Mackay and Weatherley (1973).} But trees clearly have to have strategies for recovery from cavitation when the accumulation of embolisms becomes so great as to threaten the viability of transpiration and carbon fixation -- as in temperate trees at the end of winter, before the formation of new leaves. There are in fact several such strategies; some trees use more than one at different times of the year. The most widespread  one is the annual growth of new xylem conduits. But this strategy is not available to all plants: for example, palm trees, once formed, do not grow new vascular tissue. Happily, in these trees, the embolisms appear to be restricted to leaves during times of drought, and leaves are renewable.\footnote{See Cruiziat et al (2002), p. 736} The remaining strategy is that of refilling embolized conduits, and woody plants have evolved several ways of doing this.

One such mechanism that has already been mentioned is positive root pressure, as in the case of vines and a number of trees. In the 1990s there emerged evidence of another, more mysterious, diurnal embolism-followed-by-refill mechanism in certain tree species that seems to take place  when transpiration is happening and hence when there is tension in the transpiration flow. This involves active, local refilling of an embolized conduit, where the pressurization is somehow limited to that conduit or some adjacent cells. None of the several hypotheses in the literature as to the nature of the refilling process involved has been substantiated, though it has generally been thought that the source of water that is needed in any such mechanism is the phloem.\footnote{For a brief review of this mechanism, see Wheeler and Holbrook (2007); for more detail see also Cruiziat \textit{et al.} (2002) and especially Holbrook \textit{et al.} (2002).} Despite this element of obscurity, a ``new paradigm of xylem as highly dynamic, experiencing cycles of embolism and repair has become increasingly accepted, with the ability of plants to refill under tension considered an ecologically important behaviour that allows plants to operate near their hydraulic limits.''\footnote{Wheeler \textit{et al.} (2013), p. 1938.} It is noteworthy then, and a salutary reminder of the complexity of our subject, that in 2013 it was demonstrated that at least in some tree species this ``miraculous'' refilling process is an experimental artefact\footnote{Wheeler \textit{et al.} (2013)} -- one actually anticipated by Henry Dixon.\footnote{Dixon (1914), p. 94; see Wheeler \textit{et al.} (2013), p. 1945.} It is still not known how many vulnerability-to-cavitation studies have been contaminated in this way, and there are calls for a thorough review of putative experimental evidence for the post-1990s paradigm.\footnote{See for example Sperry (2013) and Cochard and Delzon (2013). This development came too late to be incorporated in Brown (2013).} Such a review might have serious implications for the modelling of drought resistance and forest dieback in the context of climate change.\footnote{Cochard and  Delzon (2013).} 

There is, finally, another process which has particular relevance to our concerns. This has to do with sugar maple trees (\textit{Acer saccharum}). As any true aficionado of maple syrup will know, in early spring, before the appearance of leaves, a sequence of freezing nights and warm days can occur during which time sugar maple exudes sugar-rich sap from the xylem when tapped.\footnote{The sucrose in the sap is produced in the leaves and descends through the phloem, but it is transported radially into the xylem. The sugar concentration of the xylem sap is one fortieth of commercial maple syrup, obtained by boiling the exuded sap.} (The process is not dependent on root pressure, since it has been shown to occur in trunks severed from the roots but in contact with a source of water.\footnote{Stevens and Eggert (1945).}) In freezing nighttime conditions, water from the inner layers of the xylem is drawn towards ice crystals in the outer layers of the xylem and phloem, creating tension which draws sap from the roots to the upper parts of the tree -- a process only stopped by freezing. 
The uptake is greater the slower the temperature drop. Once the ice crystals thaw during the day, sap flows out, or exudes, through wounds or any convenient openings, a result of the decompression of gas trapped in the fibres during freezing. Some of the sap exuded lower in the tree has fallen from the crown of the tree, after sap was sucked up all the way up during the freeze. When exudation of maple sap is prevented during the thaw, positive pressures within the xylem can exist for many hours.

However, despite much effort on the part of plant scientists, the mechanism in sugar maple is still somewhat controversial; in particular the reason for the positive correlation between stem pressure and concentration of sucrose in the sap is still open to discussion.\footnote{For details of the two main competing models, and recent anatomical evidence in favour of the 1995 osmotic model due to Melvin Tyree, see Cirelli \textit{et al.} (2008).} For our purposes, what is of relevance is the fact that at least for certain, rather special trees, which can grow up to 45m in height, Nature has found another way of raising water from the roots -- and cells in the trunk -- to the leafless canopy. The process requires a certain diurnal variation of temperature in early spring and it depends on the cohesion of water, but neither root pressure nor of course transpiration are key elements in the process. Indeed, in the absence of this process, the start of transpiration once the leaves emerge in late spring would be in peril.

\section{Evidence}

\subsection{Cavitation again}

In his detailed 1981 review of the CT theory, William F. Pickard wrote: ``There is a great deal of evidence which is strongly supportive of cohesion-tension, some which finds no easy explanation within it, and none which decisively contradicts it. In this attribute the cohesion-tension theory is unique. It therefore is the accepted theory.''\footnote{Pickard (1981), p. 223.} Amongst the supportive evidence, Pickard cited two items. The acoustic phenomena, mentioned above, which are thought to accompany cavitations, appear to occur precisely when according to the CT theory large tensions are expected in the xylem. Furthermore, there is considerable evidence that stem diameters in trees fluctuate diurnally, and that shrinkage of stem diameter follows, with a small delay, changes in ambient conditions that are conducive to an increase in transpiration. (Further evidence is found in the 1996 review by Milburn.\footnote{Milburn (1996).}) But Pickard also raised concerns about the lack of understanding in 1981 of the process of cavitation and its repair. 

A more pressing threat to the viability of the CT theory was to arise out of the experimental work of Ulrich Zimmermann (the main author of the skeptical review paper mentioned at the start of this study) and his collaborators, starting in the 1990s. The problem was that attempts by these scientists to measure xylem pressure using intrusive pressure probes, rather than the more traditional pressure chamber (``pressure bomb'') technique, failed to register tensions as large as those predicted in the CT theory.\footnote{See Zimmermann \textit{et al.} (1994, 1995).} 

Mention was made above of the Renner-Zimmermann air-seeding mechanism for cavitation in dehydrated xylem vessels, whose widespread acceptance took place well after Pickard's 1981 review. According to this picture, when the difference between the gas pressure (normally atmospheric pressure) in an air-filled vessel and the lower pressure in an adjacent water-filled vessel reaches a critical value for a given plant species, air is pulled into the water-filled conduit through pores in the membranes of pits which link the conduits.\footnote{What prevents this seeding from taking place at typical negative xylem pressures is that the pores in the vessel walls are very small, of nanometer scale. Any curved air-water interfaces therein require, because of the large surface tension of water, a large input of energy to expand. See Pickard (1981) (who dismissed the possibility of air-seeding) and Holbrook \textit{et al.} (2002).} Part of the evidence for such a mechanism are demonstrations that the same critical pressure differences leading to cavitation obtain  in the `opposite' situation in which xylem pressure is atmospheric, but  gas pressures are artificially increased by way of introduction of compressed air. Indeed the strength of the evidence allows plant scientists to measure xylem pressure in an indirect way. If air-seeding is indeed occurring, then knowing both the critical pressure difference and the air pressure allows one to infer xylem pressure in the case of dehydrated stems. Such inferred pressures are then compared to those obtained in more direct measurements using the pressure chamber technique, and the closeness of fit is striking. It is hard to avoid the conclusion first, that the current detailed understanding of cavitation in trees due to dehydration is not only consistent with the existence of large negative (lower than atmospheric) xylem pressures, it demands it, and second that the pressure chamber technique for measuring xylem tension is essentially sound.\footnote{See Sperry (1995) and Sperry \textit{et al.} (1996). It is worth pointing out that the pressure chamber is normally attributed to 1965 work by Scholander and colleagues in the United States, but a precursor was developed forty years earlier by Dixon. According to Jones (1992), ``It can be claimed that the development of plant water relations was delayed for more than forty years by the failure of other scientists to appreciate the significance of Dixon's construction of a pressure chamber in 1922.'' (But see Tyree and Zimmermann (2002), pp. 139-140, for details of cases where the technique may lead to erroneous results.)} This conclusion is backed up by an independent centrifugal method for measuring negative xylem pressure.\footnote{Pockman \textit{et al.} (1995).}

The anomalous pressure probe results have attracted a great deal of attention, and a number of different technical suggestions -- possibly too many for comfort -- have been made as to why this procedure, which involves actually puncturing the xylem conduit, is problematic.\footnote{See in this connection Milburn (1996), Sperry \textit{et al.} (1996), Tyree (1997), Wei \textit{et al.} (1999) and the reply to the last paper by Zimmermann \textit{et al.} (2000); a detailed analysis of the pressure probe tool is given in Tomos and Leigh (1999).} It may well be that the last word has not been spoken on this matter, but what does seem clear is that the large majority of plant scientists has not been won over by Zimmermann's assault on the CT theory. One reason is surely that alternative mechanisms for transpiration-induced ascent of sap look piece-meal and implausible, if not \textit{ad hoc}.\footnote{It is worth noting that Canny (1995) proposed a vitalistic mechanism based on tissue pressure (which originates from metabolic processes) supports the ascent of sap in the xylem and plays an important role during the refilling of cavitated conduits. The theory is now widely rejected on the alleged grounds that it is inconsistent with the laws of thermodynamics; see for example Comstock (1999).} This issue would make an interesting case study for a budding philosopher of science interested in the meaning of `refutation'.\footnote{Students of the history of the special theory of relativity might recognise a certain similarity with the episode of Kauffmann's experimental `refutation' of the Lorentz-Einstein theory in experiments performed between 1901 and 1905; for details see Brown (2007), pp. 86-87.}

\subsection{Hydraulic architecture}

In 1948, building on earlier work by Hans Gradmann and others, Taco Hajo Van den Honert was to publish a seminal paper on water transport through the soil-plant-atmosphere continuum based on an analogy with Ohm's law for electrical currents.\footnote{Van den Honert (1948).} Ohm's law can be applied to any physical current that is determined by a potential difference, as is the case, to a reasonable degree of approximation, in water flow through the xylem capillaries and in a diffusion process. The language in this approach is that of resistances, capacitances and water potentials, and Van den Honert's paper was to lead to several decades of phenomenological study of water transport, where the mechanical details of the CT theory were presupposed, but not needed. In a sense, it was the opposite of the story of thermal physics, in which phenomenological thermodynamics largely predated the rise of the kinetic theory of gases and statistical mechanics. The detailed mechanism of the rise of sap through transpiration and the cohesion of water related to hydrogen bonds between molecules was replaced by a theory involving bulk properties of flow that made no commitment to the underlying molecular mechanisms. In the language of Einstein, a ``constructive'' theory was being replaced by a ``principle'' theory.\footnote{For the meaning of these terms, and the nature of thermodynamics and special relativity as principle theories, see Brown (2007), Ch 5.} Significant progress was made in determining the relative resistances of elements within the soil-tree-continuum, and the nature of the regulatory role of stomata in the leaves. But the price that is paid in this approach, as with all principle theories, is a certain loss of explanatory depth. In reading common statements such as ``evaporation from leaves lowers their water potential and causes water to move from the xylem to evaporating cells across leaf tissue''  or  ``gradients in water potential are established along transpiring plants; this causes an inflow of water from the soil into the roots and to the transpiring surfaces in the leaves'', one must not overlook the fact that flow of water is closer to being postulated, or claimed on empirical grounds, than to being explained. 

In the decades following Van den Honert's paper, most research on water transport in plants was devoted to evaluating gaseous and liquid resistances and their dependencies on climate and soil conditions, and little effort was given to developing the CT theory. This was to change in the late sixties and seventies largely through the pioneering work of J. A. Milburn, M. H. Zimmermann and M. T. Tyree. A new paradigm of ``hydraulic architecture'', a term coined by Martin Zimmermann, has gained strength in recent decades, one which attempts to combine the insights of both the Ohm's Law analogy in the soil-plant-atmosphere continuum and the CT theory. Much of the understanding of cavitation and drought-induced embolism, mentioned in the previous section, was obtained within this framework, where the emphasis is more tree-, or tree-species-, specific than in the early work on the CT theory. In particular, considerable progress has been made in the study of the variation in vulnerability within tree species, or within organs in the same species, to cavitation due to drought or freezing, with evidence emerging that root xylem -- fortunately replaceable -- is the most vulnerable part of trees. One major surprise has been the discovery that no correlation exists between the diameter of xylem conduits and vulnerability to summer embolism; on the other hand the larger the conduit, the higher the vulnerability to frost. And a much more sophisticated picture has emerged of the key role that closure of the stomata in the leaves plays in protecting against xylem dysfunction\footnote{A review of the hydraulics of leaves, which as mentioned above represent an important hydraulic bottleneck in trees, is found in Sack and Holbrook (2006). The first detailed examination of the hydraulic consequences and implications of key leaf venation traits for the economics, ecology, and evolution of plant transport capacity is McKown \textit{et al.} (2010).}, and other physiological processes such as the effect of ion concentrations on xylem hydraulic properties\footnote{See Holbrook \textit{et al.} (2002) and further references therein.}.  But these processes are still far from fully understood; in particular the molecular mechanisms which account for the fact that the stomata sense and regulate transpiration are unknown, a problem compounded by uncertainties, mentioned earlier, concerning where evaporation is occurring in the leaf.

This brief summary cannot do justice to the range of issues tackled in the hydraulic architecture approach.\footnote{For reviews see Tyree and Ewart (1991) and Cruiziat \textit{et al.} (2002).} Importantly, it has led to a resurgence of interest in the CT theory in recent decades; the theory is widely seen as more ramified and better established than ever. But there remains one development that is worth special attention in the context of our discussion of the different ways capillary forces have been harnessed by trees.

\subsection{Capillary storage}

While the large majority of water emerging from the roots goes into the xylem and evaporates from the leaves (the remainder either splitting in the photosynthetic reaction in the leaves to release O$_2$, or recirculating within the plant through the phloem, or going irreversibly into cells by growth), there is considerable evidence suggesting that in many trees the transpiration stream does not wholly originate in the roots. An important insight that evaded the pioneers of the CT theory is the ``elasticity'' of the hydraulic mechanism provided by water storage within trees.  In tall trees, the daily onset of transpiration withdraws water from internal storage compartments, with the consequence that time lags occur between changes in transpiration and the flow of sap at the base of the trees. This mechanism compensates the increase of hydraulic resistance with increasing hydraulic path length, and so helps maintain photosynthesis in the leaves of tall trees. In the case of tropical forest canopy trees, for example, and it has become apparent that stomata in the leaves play a regulatory role in limiting the use of such stored ``free'' water so that the fraction of total daily use (roughly 10\%) is independent of species and tree size.\footnote{See Meinzer (2004).} In fact, there is evidence of a striking convergence across broad-leaved trees and conifers, suggesting that water storage in sapwood (capacitance) plays an important role in tree hydraulics generally. In conifers in particular, there appear to be repeated exchanges of water between storage compartments and the transpiration stream.\footnote{Meinzer \textit{et al.} (2006).} In the case of tropical trees, it seems that water storage compartments are replenished at night, during periods of low transpiration.

 It was Martin Zimmermann who shortly before his death in 1984 had an insight as to why in certain diffuse-porous trees, like birch and maple, the water content in their stems seems to increase significantly in autumn after the fall of leaves -- a process which clearly does not depend on transpiration. Zimmermann's notion was that it was precisely the cessation of transpiration, and hence the decrease in xylem tension, that was the cause of the phenomenon. Today it is thought that several processes are at work here, but the mechanism Zimmermann anticipated is capillary storage. Water finds its way through capillary action into the tapered tips of embolized wood fibers and tracheids in the xylem; because of the laws of capillarity the amount of such trapped water varies inversely with the xylem tension, so it increases sensibly in the autumn.\footnote{For more details, see Tyree and Zimmermann (2002), section 4.8. A review of stem water storage in plants is found in Holbrook (1995).} This extra water is drawn from surrounding tissue and ultimately the roots. What is striking about this process is that it mimics the transpirational flow, but without transpiration. Water is being drawn up into the tree because of capillarity and cohesion, but the capillarity is now not situated in the leaves but within the sapwood. Another little victory for Stephen Hales?
  
\section{Final remarks}
For better of for worse, the mathematics of capillarity, diffusion, cavitation and fluid dynamics generally have been omitted, and technical details kept to a minimum, in this essay on the CT theory.\footnote{A relatively simple but illuminating analysis of the fluid dynamics of water being lifted in rigid airtight pipes with circular cross sections, involving both single and branching conduits, has been given in Denny (2012) with a view to understanding the rise of sap.} The focus of concern has been the early history of the theory, and its fundamental conceptual ingredients. The rise of sap in trees is a hugely multifaceted process; even the great Martin Zimmermann emphasised in the introduction to his 1983 textbook on the subject that he could not be an expert in all the fields involved. What this essay attempts to highlight is the simple fact that for transpiration to act as the principal engine of the upward flow of sap in trees and vines, active xylem conduits must already be water-filled, and that the process of cell growth, and sometimes root pressure,  freeze-thaw cycles, and capillarity in the sapwood play an important, often understated, role in bringing this condition about. Some of the originators of the CT theory were clearly aware of this issue, though they did not anticipate the complexity of the contributing factors.

\section{Acknowledgments}

Detailed commons on the first draft of this paper were kindly provided by Pierre Cruiziat and especially by John Sperry and members of his research group: Duncan Smith, David Love and Allison Thompson. These comments, much appreciated, corrected some errors and provided helpful suggestions for improvements. Melvin Tyree also kindly provided helpful remarks. Roger H. Stuewer gave generous assistance with bibliographical details. Above all, thanks are due to John Pannell, who provided constant encouragement, guidance on key points, and special help with understanding B\"{o}hm's 1893 paper.  Finally, this paper is dedicated to Rom Harr\'{e}, a mentor and philosopher whose extraordinarily wide interests include plant science and its history.

\section{References}

Angeles G.; Bond B.; Boyer J.S.; Brodribb T.; Brooks J.R.;
Burns M.J.; Cavender-Bares J.; Clearwater M.;
Cochard H.; Comstock J.; Davis S.D.; Domec J.-C.;
Donovan L.; Ewers F.; Gartner B.; Hacke U.; Hinckley T.; Holbrook N.M.; Jones H.G.; Kavanagh K.; Law B.; L—pez-Portillo J.; Lovisolo C.; Martin T.; Mart'nez-Vilalta J.; Mayr S.; Meinzer F.C.; Melcher P.; Mencuccini M.;
Mulkey S.; Nardini A.; Neufeld H.S.; Passioura J.; Pockman W.T.; Pratt R.B.; Rambal S.; Richter H.; Sack L.; Salleo S.; Schubert A.; Schulte P.; Sparks J.P.; Sperry J.; Teskey R.; Tyree M., ``The Cohesion-Tension Theory'', \textit{New Phytologist} \textbf{163} (2004), 451Ð 452.

Arora, Vivek K. and Montenegro, Alvaro, ``Small temperature benefits provided by realistic afforestation efforts'', \textit{Nature Geoscience} \textbf{4} (2011), 514-518.

Askenasy, Eugen, ``\"{U}ber das Saftsteigen'', \textit{Verhandlungen des Naturhistorisch-Medizinischen Vereins zu Heidelberg} \textbf{5} (1895), 325-345; \textit{idem},
``Beitrage zur ErklŠrung des Saftsteigens'', \textit{ibid} \textbf{5} (1896), 429-448. 

Bala, G.;  Caldeira, K.; Wickett, M.;, T. J. Phillips, T.J.; D. B. Lobell, D.B.; C. Delire, C. and A. Mirin, A., ``Combined climate and carbon-cycle effects of large-scale deforestation'', \textit{Proceedings of the National Academy of Sciences} \textbf{104} (2007), 6550-6555.

Beerling, David J, and Franks, Peter J., ``The hidden cost of transpiration'', \textit{Nature} \textbf{465} (2010), 495-496.

Betts, Richard A., ``Afforestation cools more or less'', \textit{Nature Geoscience} \textbf{4} (2011), 504-505.

B\"{o}hm, Josef, ``Capillarit\"{a}t und Saftsteigen'', \textit{Bericht der Deutschen botanischen Gesellschaft} \textbf{11} (1893), 203-212.

Bonan, George B., ``Forests and Climate Change: Forcings, Feedbacks, and the Climate Benefits of Forests''
\textit{Science} \textbf{320} (2008), 1444-1449.

Bourne, J., ``Redwoods. The Super Trees'', \textit{National Geographic} (October 2009).

Boyer, J.S., ``Water transport'', \textit{Annual Review of Plant Physiology} \textbf{36} (1985), 473-516.

Boyer, J. S.,``Cell enlargement and growth-induced water potentials''. \textit{Physiologia Plantarum} \textbf{73} (1988), 311-316.

Boyer, J.S., Cavalieri, A.J., and Schulze, E.D. (1985) ``Control of the rate of cell enlargement: excision, wall relaxation, and growth-induced water potentials.'' \textit{Planta} \textbf{163} (1985), 527Ð543

Brown, Harvey R., \textit{Physical Relativity. Space-time structure from a Dynamical Perspective}, (Oxford: Clarendon Press, 2007).

Brown, Harvey R., ``The theory of the rise of sap in trees: Some historical and conceptual remarks'', \textit{Physics in Perspective} \textbf{15} (2013), 320-358.

Canny, M. J., ``Flow and Transport in Plants'', \textit{Annual Review of Fluid Mechanics} \textbf{9} (1977), 275-296.

Canny, M. J., ``A new theory for the ascent of sap -- cohesion supported by tissue pressure'', \textit{Annals of Botany} \textbf{75} (1995), 343-357.

Carder, Al C. (1995). \textit{Forest Giants of the World, Past and Present} (Markham, Ontario : Fitzhenry and Whiteside, 1995).

Chang, Hasok, \textit{Inventing Temperature. Measurement and Scientific Progress} (Oxford: Oxford University Press, 2007).

Cirelli, Damien; Jagels, Richard and Tyree, Melvin T., ``Toward an improved model of maple sap exudation: the location and role of osmotic barriers in sugar maple, butternut and white birch'', \textit{Tree Physiology} \textbf{28} (2008), 1145-1155.

Cochard, Herv\'{e} and Delzson, Sylvain, ``Hydraulic failure and repair are not routine in trees'', \textit{Annals of Forest Science} \textbf{70} (2013), 659-661.

Comstock, Jonathan P., ``Why Canny's theory doesn't hold water'', \textit{American Journal of Botany} \textbf{86} (1999), 1077-1081.

Copeland, Edwin B., ``The Rise of the Transpiration Stream: An Historical and Critical Discussion'', \textit{Botanical Gazette} \textbf{34} (1902), 161-193, 260-268. 

Cruiziat, Pierre, and Richter, Hanno,``The CohesionÐTension Theory at Work'', Essay 4.2 (2006), in \textit{Plant Physiology Online, Fifth Edition}; \newline http://5e.plantphys.net/article.php?ch=4\&id=99.

Cruiziat, Pierre; Cochard, Herv\'{e} and Am\'{e}glib, Thierry, ``Hydraulic architecture of trees: main concepts and results'', \textit{Annals of Forrest Science} \textbf{59} (2002), 723-752.

Darwin, Francis; Vines, S.H.; Joly, J. and FitzGerald, G.F., ``Report of a discussion on the ascent of water in trees'' \textit{Annals of Botany (London)} \textbf{10} (1896), 630-661.

Denny, Mark, ``Tree hydraulics: how sap rises'', \textit{European Journal of Physics} \textbf(33) (2012), 45-53.

Dixon, Henry H., ``Transpiration and the ascent of sap in plants'', (London: Macmillan, 1914).

Dixon, Henry H., and Joly, John, ``On the Ascent of Sap (Abstract)'', \textit{Proceedings of the Royal Society of London} \textbf{57} (1894), 3-5.

Dixon, Henry H., and Joly, John, ``On the Ascent of Sap'', \textit{Philosophical Transactions of the Royal Society of London B} \textbf{186} (1895), 563-576.

Donny, M. F., ``XLVIII. On the cohesion of liquids and their adhesion to solid bodies'', \textit{Philosophical Magazine Series 3}, \textbf{28: 187} (1846), 291-294.

Ehlers, Wilfred and Goss, Michael, \textit{Water dynamics in plant production} (Wallingford: CABI Publishing, 2003).

Eilperin, Juliet, ``Protecting pine forests from warming threat'', \textit{The Washington Post}, May 14, 2012.

Floto, Franz, ``Stephen Hales and the cohesion theory'', \textit{Trends in Plant Science} \textbf{4} (1999), 209.

Gartner, Barbara L. (ed.), \textit{Plant Stems: Physiology and Functional Morphology} (London: Academic Press, 1995).

Gerrienne, Philippe; Gensel, Patricia G.; Strullu-Derrien, Christine; Lardeux, Hubert; Steemans, Philippe; Prestianni, Cyrille, ``A Simple Type of Wood in Two Early Devonian Plants'' \textit{Science} \textbf{333} (2011), 837.

Greenidge, K.N.H., ``Ascent of sap'', \textit{Annual Review of Plant Physiology} \textbf{8} (1957), 237-256.

Hales, Stephen, \textit{Vegetable Staticks: Or, An Account of some Statical Experiments on the Sap in VEGETABLES}, etc. (London: W. and J. Innys and T. Woodward, 1727; London. Reprint 1969 London: MacDonald  and New York: American Elsevier 1969).

Harr\'{e}, Rom. \textit{The Method of Science. A course in Understanding Science, based upon the} De Magnete \textit{of William Gilbert and the} Vegetable Staticks \textit{of Stephen Hales} (London and Winchester: Wykeham Publications Ltd., and New York, Springer-Verlag, 1970).

Hartley, Iain P.; Garnett, Mark H.; Sommerkorn, Martin; Hopkins, David W.; Fletcher, Benjamin J.;, Sloan, Victoria L.;, Phoenix, Gareth K. and Wookey, Philip A., ``A potential loss of carbon associated with greater plant growth in the European Arctic''\textit{Nature Climate Change}, published online 17 June 2012.

Helmholtz, Herman, ``Ueber galvanische Polarisation in gasfreien Fl\"{u}ssigkeiten'' [1873], in \textit{Wissenschaftliche Abhandlungen}. Erster Band (Leipzig: Johann Ambrosius Barth, 1882), pp. 821-834.

Holbrook, N. Michele, ``Stem Water Storage'', in Gartner (1995), 151-174.

Holbrook, N. Michele; Zwieniecki, Maclej A. and Melcher, Peter J., ``The Dynamics of ``Dead Wood": Maintenance of Water Transport through Plant Stems'' \textit{Integrative and Comparative Biology} \textbf{42} (2002), 492-496.

Jones, Michael, `` `A name write in water': Henry Horatio Dixon 1869-1953'', in \textit{Treasures of the Mind.  A Trinity College Dublin Quatercentenary Exhibition}, David Scott (ed.) (London: Sotheby's, 1992), pp. 97-103. Online version at http://www.tcd.ie/Botany/tercentenary/300-years/chairs/henry-horatio-dixon.php.

Kell, George S., ``Early observations of negative pressures in liquids'', \textit{American Journal of Physics} \textbf{51} (1983), 1038-1041.

Koch, George; Stephen Sillett, Stephen; Jennings, Gregg; and Davis, Stephen, ``How Water Climbs to the Top of a 112 Meter-Tall Tree'', Essay 4.3 (2006), in \textit{Plant Physiology Online, Fifth Edition}; \newline http://5e.plantphys.net/article.php?ch=4\&id=100.

Kozlowski, T.T. and Pallardy, S.G., \textit{Physiology of Woody Plants} (San Diego: Academic Press, 1996).

Kramer, Paul K. and Boyer, John S., \textit{Water Relations of Plants and Soils} (San Diego, London: Academic Press, 1995).

Lindenmayer, David B.; Laurance, William F. and Franklin, Jerry F., ``Global decline in large old trees'',  \textit{Science} \textbf{338} (2010), 1035-1036.

Mackay, J.F.G. and Weatherley, P.E., ``The effects of transverse cuts through the stems of transpiring woody plants on water transport and stress in the leaves'', \textit{Journal of Experimental Botany} \textbf{24} (1973), 15Ð28.

McKown, Athena D.; Cochard, Herv\'{e} and Sack, Lawren, ``Decoding Leaf Hydraulics with a Spatially Explicit Model: Principles of Venation Architecture and Implications for Its Evolution'',  \textit{The American Naturalist} \textbf{175} (2010), 447-460.

Meinzer, Frederick C.; James, Shelley A. and Goldstein, Guillermo, ``Dynamics of transpiration, sap flow and use of stored water in tropical forest canopy trees'', \textit{Tree Physiology} \textbf{24} (2004), 901-909.

Meinzer, Frederick C.; Brooks, F.C.; Domec, J.-C.; Gartner, B.L.; Warren, J.M., Woodruff, D.R.; Bible K. and Shaw D.C., ``Dynamics of water transport and storage in conifers studied with deuterium and heat tracing techniques''  \textit{Plant, Cell and Environment} \textbf{29} (2006), 105-114.

Meylan, B. A. and Butterfield B. G., \textit{Three dimensional structure of wood} (Hong Kong: Chapman and Hall, Ltd., 1972).

Milburn, John A., ``Sap Ascent in Vascular Plants: Challengers to the Cohesion Theory Ignore the Significance of Immature Xylem and the Recycling of M\"{u}nch Water'', \textit{Annals of Botany} \textbf{78} (1996), 399-407.

Miller, Edwin C., \textit{Plant Physiology; with reference to the green plant} (New York and London: McGraw-Hill Book Company, Inc., 1938).

Newton, Isaac, \textit{Opticks: or a treatise on the reflexions, refractions, inflexions and colours of light}, Based on fourth edition London 1730 (New York, Dover Publications, 1979).

Nobel, Park S. \textit{Physicochemical and Environmental Plant Physiology (Third Edition)} (Elsevier, 2005). 

Oertli, J. J.,``The stability of water under tension in the xylem'', \textit{Zeitschrift f\"{u}r Pflanzenphysiologie} \textbf{65} (1971), 195-209.

Paasonen, P.; Asmi, A.; Pet\"{a}j\"{a}, T.; Kajos M.K.; \"{A}ij\"{a}l\"{a}, M.; Junninen, H.; Holst, T.; Abbatt, J.P.D.; Arneth, A.; Birmili, W.; van der Gon, H.D.; Hamed, A.; Hoffer, A.; Laakso, L.;  Laaksonen, A.; Leaitch, W.R.;
Plass-D\"{u}lmer, C.; Pryor, S.C.; R\"{a}is\"{a}nen, P.; Swietlicki, E.; Wiedensohler, A.; Worsnop, D.R.; Kerminen, V.-M. and Kulmala, M., ``Warming-induced increase in aerosol number concentration likely to moderate climate change'' \textit{Nature Geoscience} \textbf{6} (2013) 438-442.

Pickard, William F., ``The Ascent of Sap in Plants'', \textit{Progress in Biophysics \& Molecular Biology} \textbf{37} (1981), 181-229.

Pickard, William F., ``How might a treachery element which is embolized by day be healed by night?'', \textit{Journal of Theoretical Biology} \textbf{141} (1989), 259-279.

Pockman, William T.; Sperry, John S. and O'Leary, James W., ``Sustained and significant negative water pressure in xylem'', \textit{Nature} \textbf{378} (1995), 715-716.

J. Pongratz, J.; Reick, C.H.; Raddatz, T.; Caldeira,K. and Claussen, M.,``Past land use decisions have increased mitigation potential of reforestation'', \textit{Geophysical Research Letters} \textbf{38} (2011), L15701.

Quinton, Sophie, ``As politicians debate climate change, our forests wither'', http://www.theatlantic.com/national/archive/2012/06/as-politicians-debate-climate-change-our-forests-wither/258549/.

Richter, Hanno and Cruiziat, Pierre, ``A Brief History of the Study of Water Movement in the Xylem'', Essay 4.1
(2002), in \textit{Plant Physiology Online, Fifth Edition}; http://5e.plantphys.net/article.php?ch=\&id=98

Rowlinson, J.S., \textit{Cohesion. A Scientific History of Intermolecular Forces} (New York: Cambridge University Press, 2005).

Sack, Lauren and Holbrook, N. Michele, ``Leaf Hydraulics'', \textit{Annual Review of Plant Biolology} \textbf{57}(2006), 361-381.

Scholander, P. F.; Love, Warner E. and Kanwisher, John W., ``The Rise of Sap in Tall Grapevines'', \textit{Plant Physiology} \textbf{30} (1955), 93-104.

Shmulsky, Rubin and Jones, P. David, \textit{Forest Products and Wood Science: and Introduction, Sixth Edition}.  (John Wiley \& Sons, Inc, 2011).

Smith, A.M., ``Negative pressure generated by octopus suckers: a study of the tensile strength of water in nature'', \textit{Journal of Experimental Biology} \textbf{157} (1991), 257-271.

Sperry, John S., ``Limitations on Stem Water Transport and Their Consequences'', in Gartner (1995), 105-124.

Sperry, John S., ``Hydraulics of vascular water transport'',  in: Wojtaszek P. (ed.), \textit{Mechanical Integration of Plant Cells and Plants  [Signaling and Communication in Plants} Vol. 9 (Berlin and Heidelberg, Springer-Verlag, 2011); pp. 303-327.

Sperry, John S., ``Cutting-edge research or cutting-edge artefact? An overdue control experiment complicates the xylem refilling story'', \textit{Plant, Cell and Environment} \textbf{36} (2013), 1916-1918.

Sperry, John S.; Saliendra, N.Z.; Pockman, W.T.; Cochard, H.; Cruiziat, P.;  Davis, S.D.; Ewers, F.W.  and M.T. Tyree, M.T., ``New evidence for large negative xylem pressures and their measurement by the pressure chamber method'', \textit{Plant, Cell and Environment''} \textbf{19}  (1996), 427-436.

Spracklen D.V.;  Bonn, B. and Carslaw, K.S., ``Boreal forests, aerosols and the impacts on clouds and climate'', \textit{Philosophical Transactions of the Royal Society A} \textbf{366} (2008), 4613-4626.

Steudle, Ernst, ``Trees under tension'', \textit{Nature} \textbf{378} (1995), 663-664.

Steudle, Ernst, ``The Cohesion-Tension Mechanism and the Acquisition of Water by Plant Roots'', \textit{Annual Review of Plant Physiology and Plant Molecular Biololgy} \textbf{52} (2001), 847-875.

C L Stevens, C.L. and R L Eggert, R.L., ``Observations on the causes of the flow of sap in red maple'', \textit{Plant Physiology} \textbf{20} (1945), 636-648.

Strasburger, E., \textit{Ueber den Bau und die Verrichtungen der Leitungsbahnen in den Pflanzen} (Jena: Gustav Fischer, 1891); \textit{idem, \"{U}ber das Saftsteigen} (Jena: Gustav Fischer, 1893).

Suzuki, David and Grady, Wayne, \textit{Tree. A Life Story} (Vancouver: Greystone Books, 2004).

Tanner, W. and Beevers, H., ``Does transpiration have an essential function in long-distance ion transport in plants?'', \textit{Plant, Cell \& Environment} \textbf{13} (1990), 745-879.

Tomos, Deri and Leigh, Roger A., ``The Pressure Probe: A Versatile Tool in Plant Cell Physiology'', \textit{Annual Review of Plant Physiology and Plant Molecular Biololgy} \textbf{50} (1999), 447-472.

Trevena, David H., ``Marcelin Berthelot's First Publication in 1850, on the Subjection of Liquids to Tension'', \textit{Annals of Science} \textbf{35} (1978), 45-54. 

Tudge, Colin, \textit{The Secret Life of Trees: How They Live and Why They Matter} (London: Penguin Press Science, 2006).

Tyree, Melvin T., ``The cohesion-tension theory of sap ascent: current controversies'',  \textit{Journal of Experimental Botany} \textbf{48} (1997), 1753-1765.

Tyree, Melvin T. and Ewers, Frank W., ``The hydraulic architecture of trees and other woody plants'', \textit{New Phitologist} \textbf{119} (1991), 345-360.

Tyree, Melvin T.; Salleo, Sebastiono; Nardini, Andrea; Lo Gullo, Maria Assunta and Mosca, Roberto,  ``Refilling of Embolized Vessels in Young Stems of Laurel. Do We Need a New Paradigm?'', \textit{Plant Physiology} \textbf{120} (1999), 11-22.

Tyree, Melvin T. and Sperry, J.S., ``Vulnerability of Xylem to Cavitation and Embolism'', \textit{Annual Review of Plant Physiology and Plant Molecular Biololgy} \textbf{40} (1989), 19-38.

Tyree, Melvin T. and Zimmermann, Martin H., \textit{Xylem Structure and the Ascent of Sap} (Heidelberg: Springer-Verlag, 2002).

Van Den Honert, T.H., ``Water transport in plants as a catenary process'', \textit{Discussions of the Faraday Society} \textbf{3} (1948), 146Ð153.

Vines, S.H., ``The Suction-force of Transpiring Branches'', \textit{Annals of Botany}, \textbf{10} (1896), 429-444.

Wheeler, James K. and Holbrook, N. Michele, ``Cavitation and Refilling'', Essay 4.4 (2007), in \textit{Plant Physiology Online, Fifth Edition}; \newline http://5e.plantphys.net/article.php?ch=4\&id=395.

Wheeler, J.K.; Huggett, B.A.; Tofte, A.N.; Rockwell, F,E.and Holbrook, N.M. (2013) ``Cutting xylem under tension or supersaturated with gas can generate PLC and the appearance of rapid recovery from embolism'', \textit{Plant, Cell and Environment} \textbf{36} (2013),1938Ð1949.

Wei, Chunfang; Steudle, Ernst and Tyree, Melvin T.,  ``Water ascent in plants: do the ongoing controversies have a sound basis?'', \textit{Trends in Plant Science} \textbf{4} (1999), 372-375.

Wyse Jackson, Patrick N., ``A man of invention: John Joly (1857-1933), Engineer, Physicist and Geologist'', in \textit{Treasures of the Mind.  A Trinity College Dublin Quatercentenary Exhibition}, David Scott (ed.) (London: Sotheby's, 1992), pp. 89-96.

Zhao, Maosheng and Running, Steven W.,  ``Drought-Induced Reduction in Global Terrestrial Net Primary Production from 2000 Through 2009'', \textit{Science} \textbf{329} (2010), 940-943.

Zimmermann, Martin H.,``How sap moves in trees'', \textit{Scientific American} \textbf{208} (March 1963a), 132-142.

Zimmermann, Martin H., \textit{Xylem structure and the ascent of sap}, (Berlin: Springer Verlag, 1983b).

Zimmermann, Ulrich; Meinzer, F.; and Bentrup, Freidrich-Wilhem, ``How does water ascend in tall trees and other vascular plants?'', \textit{Annals of Botany} \textbf{76} (1995), 545-551.

Zimmermann, Ulrich;  Schneider, Heike; Wegner, Lars H.  and Haase, Axel, ``Water ascent in tall trees: does evolution of land plants rely on a highly metastable state?'', \textit{New Phytologist} \textbf{162} (2004), 575-615.

Zimmermann, Ulrich;  Wagner, Hans-J\"{u}rgen; Schneider, Heike; Rokitta, Markus; Haase, Axel; and Bentrup, Freidrich-Wilhem, ``Water ascent in plants: the ongoing debate'', \textit{Trends in Plant Science} \textbf{5} (2000), 145-146.

Zimmermann, Ulrich; Zhu, J.J.; Meinzer, F.; Goldstein, G.; Schneider, H.; Zimmermann, G.; Bankert, R.; Thurmer, R.; Melcher, P.; Webb, D.; and Haase, A., ``Xylem water transport : is the available evidence consistent with the cohesion theory?'',  \textit{Plant, Cell and Environment} \textbf{17} (1994), 1169-1181.

\end{document}